\documentclass[%
reprint,
amsmath,amssymb,
aps,
superscriptaddress,
prc
]{revtex4-1}

\usepackage{graphicx}
\usepackage{dcolumn}
\usepackage{bm}
\usepackage{hyperref}

\usepackage[capitalise]{cleveref}
\usepackage{comment}

\newcommand{\GeVc}          {Ge\kern-.1emV/$c$\xspace}
\newcommand{\TeV}           {Te\kern-.1emV\xspace}
\newcommand{\GeV}           {Ge\kern-.1emV\xspace}

\begin{document}
	
	
	\title{
    Exploring the multiplicity dependence of the flavor hierarchy for hadron productions in high energy pp collisions
 }
	
	\author{Aogui Zhang}
	\affiliation{School of Mathematics and Physics, China University of
		Geosciences (Wuhan), Wuhan 430074, China}
  
	\author{Xinye Peng}
	\affiliation{School of Mathematics and Physics, China University of
		Geosciences (Wuhan), Wuhan 430074, China}
	\affiliation{Key Laboratory of Quark and Lepton Physics (MOE) and Institute
		of Particle Physics, Central China Normal University, Wuhan 430079, China}
  
	\author{Liang Zheng}\email{zhengliang@cug.edu.cn}
	\affiliation{School of Mathematics and Physics, China University of
		Geosciences (Wuhan), Wuhan 430074, China}
	\affiliation{Shanghai Research Center for Theoretical Nuclear Physics, NSFC and Fudan University, Shanghai 200438, China}

	\date{\today}
	
	\begin{abstract}
    In this work, we perform a systematic study on the multiplicity dependence of hadron productions at mid-rapidity ($|y|<0.5$), ranging from the light to the charm sector in pp collisions at $\sqrt{s}=13$ TeV. This study utilizes a multi-phase transport model (AMPT) coupled with PYTHIA8 initial conditions. We have investigated the baryon to meson ratios as well as the strange to non-strange meson ratios varying with the charged particle density. By tuning the coalescence parameters, the AMPT model provides a reasonable description to the experimental data for inclusive productions of both light and charm hadrons, comparable to the string fragmentation model calculations with color reconnection effects. Additionally, we have  analyzed the relative production of hadrons by examining self-normalized particle ratios as a function of the charged hadron density. Our findings suggest that parton evolution effects and the coalescence hadronization process in AMPT model lead to a strong flavor hierarchy in the multiplicity dependence of the baryon to meson ratio. Furthermore, our investigation on the $p_T$ differential double ratio of baryon to meson fraction between high and low multiplicity events indicates distinct modifications to the flavor associated baryon to meson ratio $p_T$ shape in high multiplicity events when comparing the coalescence hadronization model to the color reconnection model. These observations highlight the importance of understanding the hadronization process in high-energy proton-proton collisions through a comprehensive multiplicity dependent multi-flavor analysis.

	\end{abstract}
	

	\maketitle
	
	
	\section{Introduction}
	
Collective phenomena have long been considered crucial signatures for the formation of a deconfined state of nuclear matter, Quark-Gluon Plasma (QGP), in high-energy heavy-ion collisions~\cite{Broniowski:2008vp,Busza:2018rrf,Elfner:2022iae,Harris:2023tti,ZHANG2023Experimental}. However, in recent years a flood of similar collectivity-like features have been witnessed in smaller systems, namely high-multiplicity proton-proton (pp) and proton-nucleus (pA) interactions at the Relativistic Heavy-Ion Collider and the Large Hadron Collider (LHC)~\cite{Adolfsson:2020dhm,CMS:2010ifv,CMS:2016fnw,ALICE:2016fzo,ALICE:2015ial,ALICE:2018pal,Shou:2024uga}. These findings, largely unanticipated for such small systems, indicate potential similarities between the collective behavior observed in small and large systems, demanding a paradigm shift in our understanding of the QGP matter~\cite{Nagle:2018nvi,Noronha:2024dtq}. Investigating the system size dependence of these collectivity phenomena can be an important way to understand the property of the deconfined quark matter created in different collision processes.

The system size of different collisions can be effectively classified based on the event multiplicity usually represented by the final state charged-particle pseudorapidity density measured at midrapidity $dN/d\eta$~\cite{ALICE:2017jyt,ALICE:2020nkc,ALICE:2023edr,ALICE:2022imr}. The variations in hadron yields related to the particle species~\cite{ALICE:2019avo,ALICE:2019etb,ALICE:2020nkc}, and characteristic shifts in $p_{T}$-dependent baryon-to-meson ratios ~\cite{ALICE:2020nkc,ALICE:2021npz,ALICE:2019avo,ALICE:2018pal} with different final state multiplicities are remarkable initial sparks indicating the existence of the collectivity effect identified in small systems. 
A continuous transition of hadron productions as a function of $dN/d\eta$ is found from pp collisions to much heavier ion collisions~\cite{ALICE:2018pal,ALICE:2016fzo}. The smooth evolution of the yield and abundances of different hadron species suggests the existence of a common underlying mechanism determining the chemical composition of particles produced in these different collision systems.

The distinguishing modifications to hadron production process such as the baryon to meson fraction changes are investigated using hadrons consisting of light flavor quarks at the beginning.
The resemblance of the light flavor hadron productions between high multiplicity pp collisions and heavy ion collisions in the soft regime stimulates the application of hydrodynamic and thermodynamic modeling to describe the bulk particle yields in small systems~\cite{Kanakubo:2018vkl,Kanakubo:2019ogh,Vislavicius:2016rwi,Zhao:2017rgg,Zhao:2020pty,Dong:2023zbu,Tang:2023wcd}. The measured relative abundances of the created particles can be used as important experimental inputs to constrain the temperature, chemical potential and volume of the produced matter in pp collisions~\cite{Mazeliauskas:2019ifr,Motornenko:2019jha,Flor:2021olm,Biro:2020kve}. 
Another phenomenological modeling approach often relies on the modified string fragmentation framework implemented based on the multi-parton interaction (MPI) assumption~\cite{Bierlich:2015rha,Bierlich:2016vgw,Bierlich:2017vhg,Bierlich:2020naj,Bierlich:2021poz,Bierlich:2024odg}. It is expected that the inter-string effects can be sizable in the dense environment with multiple MPI string system overlapped in the coordinate space. The color reconnection and rope hadronization effect implemented in the PYTHIA8 model are found to be successful in describing the multiplicity dependence of the flow-like behavior of particle spectra in pp collisions~\cite{OrtizVelasquez:2013ofg,Bierlich:2015rha,Bierlich:2014xba}.

Recently, similar multiplicity dependent measurements have been extended to the charm hadrons in high energy pp collisions, a sizable enhancement of baryon-to-meson ratio $\Lambda_c^+/D^0$ at intermediate $p_T$ is also found~\cite{ALICE:2023sgl,ALICE:2023wbx}. Unlike the light flavor hadrons, charm hadrons are usually produced through hadronization of the charm quarks originating from the initial hard scattering process. The substantial differences in charm quark fragmentation fractions observed in LHC pp collisions compared to those in electron-positron and electron-proton collisions have been a surprising revelation. The origin of these discrepancies is currently under hot debates.~\cite{Kniehl:2020szu,CMS:2023frs,ALICE:2023sgl}. Calculations based on PYTHIA8 using the fragmentation parameters tuned to electron positron and electron proton collisions can not describe this charm hadron production feature in pp collisions. However, the PYTHIA8 results with color reconnection effect included are found to provide much better agreement with the experimental data~\cite{ALICE:2023sgl,ALICE:2023wbx,Bierlich:2023okq}. Models taking into account of the coalescence mechanism for charm quark fragmentation can also reproduce the experimental data satisfactorily assuming the creation of deconfined quark matter in the small system~\cite{Song:2018tpv,Chen:2020drg,He:2019tik,Minissale:2020bif,Zhao:2023ucp}. This array of hadron production measurements across the flavor spectrum hints the necessity of modifications to the fragmentation process in vacuum. A systematic study to understand the multiplicity dependence of the flavor hierarchy can be of great interest to differentiate the underlying explanations for the flow-like effects in pp collisions.  

In this work, we employ the string melting AMPT model built on PYTHIA8 initial conditions with the final state interactions and parton coalescence mechanism included to study the multiplicity dependent hadron productions of various flavors. The AMPT model with PYTHIA8 initial conditions is found to reasonably describe the hadron yield in soft regime and the multi-partcle correlations simultaneously~\cite{Lin:2021mdn,Zheng:2021jrr}. Being capable of delivering the final state rescattering effects at both partonic and hadronic level, the AMPT model provides an important way to test the final state effects for hadron productions from light to heavy flavors in presence of the deconfined parton matter. We will compare the AMPT results to the string fragmentation model calculations in the multiplicity dependent flavor hierarchy of hadron productions to demonstrate the key features of these two widely used physics assumptions for small system collectivity studies.

The rest of this paper is organized as follows: we explain the model setups for the AMPT model and the color reconnection included string fragmentation model in Sec.~\ref{sec:formalism}. The results of the model calculations are presented and compared to the experimental data in Sec.~\ref{sec:results}. We summarize the major conclusions and discuss the future applications in Sec.~\ref{sec:summary}.

	\section{\label{sec:formalism} Method}

    In this study, we have carried out the research using the AMPT model based on PYTHIA8 initial conditions to explore the final state interaction effects on the hadron productions with different flavor components. The string melting AMPT model is consisting of four major ingredients: fluctuating initial conditions, final state parton transport interactions, the coalescence hadronization model, and final state hadronic cascade interactions. The event by event fluctuating initial conditions for the subsequent evolution stage are generated using PYTHIA8~\cite{Sjostrand:2014zea} embedded with the spatial structure at the sub-nucleon level. After propagating the initial string system to their formation time and converting them to the constituent valence quark components, the resultant quark system may experience the parton evolution stage with the microscopic scattering process implemented by the Zhang's Parton Cascade (ZPC) model~\cite{Zhang:1997ej}, with a two-body scattering cross section $\sigma$ usually determined by comparing to the anisotropic flow data. In this work, the value of this parton parton scattering cross section is set to $\sigma=0.15$ mb which gives a satisfactory description to the elliptic flow measurements in pp collisions at $\sqrt{s}=$13 TeV~\cite{Zheng:2024xyv}. When the partons cease interactions during the evolution stage, they combine with their nearby quarks using the improved spatial coalescence model~\cite{He:2017tla}. An overall coalescence parameter $r_{BM}$ has been introduced to determine the relative probability for a quark to become a meson or a baryon in this model. To describe the hadron yield for particles with different quark flavors, we follow the prescriptions in Refs.~\cite{Shao:2020sqr,Zheng:2019alz,Lin:2021mdn} and adjust the coalescence parameter for each flavor sector. For non-strange light flavor quark clusters, we have the coalescence parameter $r_{BM}=0.53$. If a strange quark or a charm quark is involved in the coalescence process, the value of this parameter has been changed to $r_{BM}^{s}=0.9$ and $r_{BM}^{c}=1.4$, respectively. As will be shown in Sec.~\ref{sec:results}, these choices can lead to a reasonable description to the inclusive productions for all the investigated hadron species. The hadrons after coalescence may undergo further hadronic rescatterings described by the extended relativistic transport model (ART)~\cite{Lin:2004en,Li:1995pra}. It must be noted that the coalescence parameters may significantly influence the magnitude of the baryon-to-meson ratio in a global way, they do not introduce strong multiplicity dependence. On the other hand, the fraction of initial strings surviving the parton evolution stage also impacts the final hadron production ratios. The hadron ratios produced from initial strings without participating the evolution in ZPC stage reflect the characteristics of the string fragmentation feature. In low-multiplicity events, hadronization from these non-interacting initial strings plays a significant role in hadron productions. Conversely, in high-multiplicity events, the coalescence process becomes the dominant factor in determining the particle ratios. This interesting feature provides valuable insights into the collision system and its hadronization mechanism.
    
    In this work, we turn on the parton and hadron final state transport mechanism in a step by step way to explore the effects developed in different evolution stages. When both parton and hadron rescattering are disabled, the results are labeled as "noFSI" (no final-state interaction), whose behavior should be similar to the pure PYTHIA string fragmentation predictions without any collective effect. If the parton rescattering stage is enabled while hadron rescatterings are excluded, it is denoted as the "pFSI" case (partonic final state interaction). When both final state parton and hadron rescattering effects are included, the results are indicated as "allFSI" (all final-state interaction), in which the evolving system experiences the entire partonic and hadronic evolutions.

    On the other hand, it has been conceived that the collectivity like behavior in small systems can be also induced by the modified string fragmentation models which take inter-string effects into consideration when a significant amount of string pieces are overlapped in the limited transverse space~\cite{OrtizVelasquez:2013ofg}. The color reconnection (CR) model has been found to reasonably reproduce the inclusive baryon to meson ratios with different flavors~\cite{Bierlich:2014xba,Bierlich:2021poz,ALICE:2023wbx,Bierlich:2023okq}. In this work, we employ the beyond leading color (BLC) CR model built in PYTHIA8.309 package~\cite{Sjostrand:2014zea}, in which strings are allowed to form between both leading and non-leading connected partons~\cite{Christiansen:2015yqa}. With the possibility to form junction in beyond leading color CR as additional source for baryon production, a multiplicity dependent baryon enhancement is observed in this model~\cite{Bierlich:2015rha}. We will compare the AMPT calculations to the results from the CR model to explore the difference between these two underlying physics mechanisms. The parameters of the CR model used in this work are set following the same procedure described in Refs.~\cite{Bierlich:2014xba,Cui:2022puv}.

\begin{figure*}[hbt!]
    \centering
\includegraphics[width=0.48\textwidth]{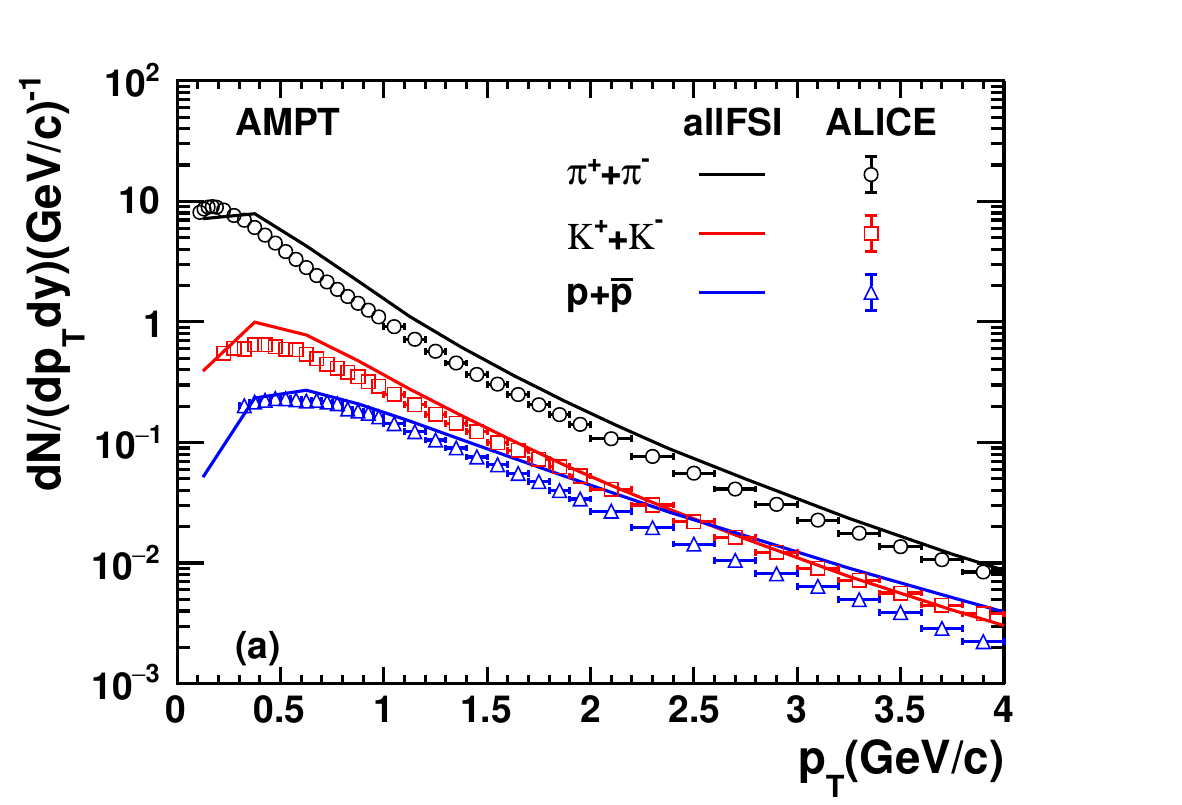}
\includegraphics[width=0.48\textwidth]{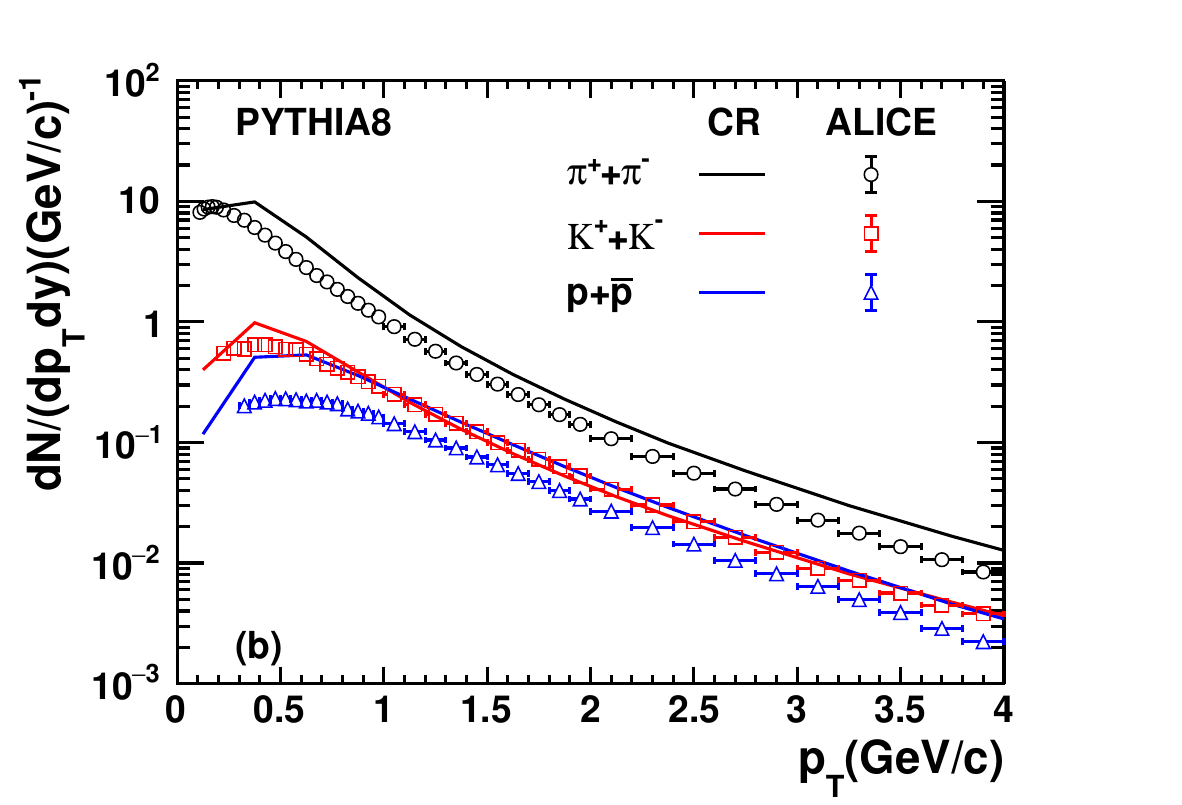}
    \caption{Transverse momentum spectra of charged $\pi^{\pm}$ (black), $K^{\pm}$ (red), and $p+\bar{p}$ (blue) in $\sqrt{s}=13$ TeV pp collisions at mid-rapidity $|y|<0.5$. The lines represent model calculations from AMPT with all final state interactions (a) and PYTHIA8 model with color reconnection effects (b), while the open markers represent ALICE data~\cite{ALICE:2020jsh}.}
    \label{fig:pt_pikp}
\end{figure*}

\begin{figure*}[hbt!]
    \centering
\includegraphics[width=0.48\textwidth]{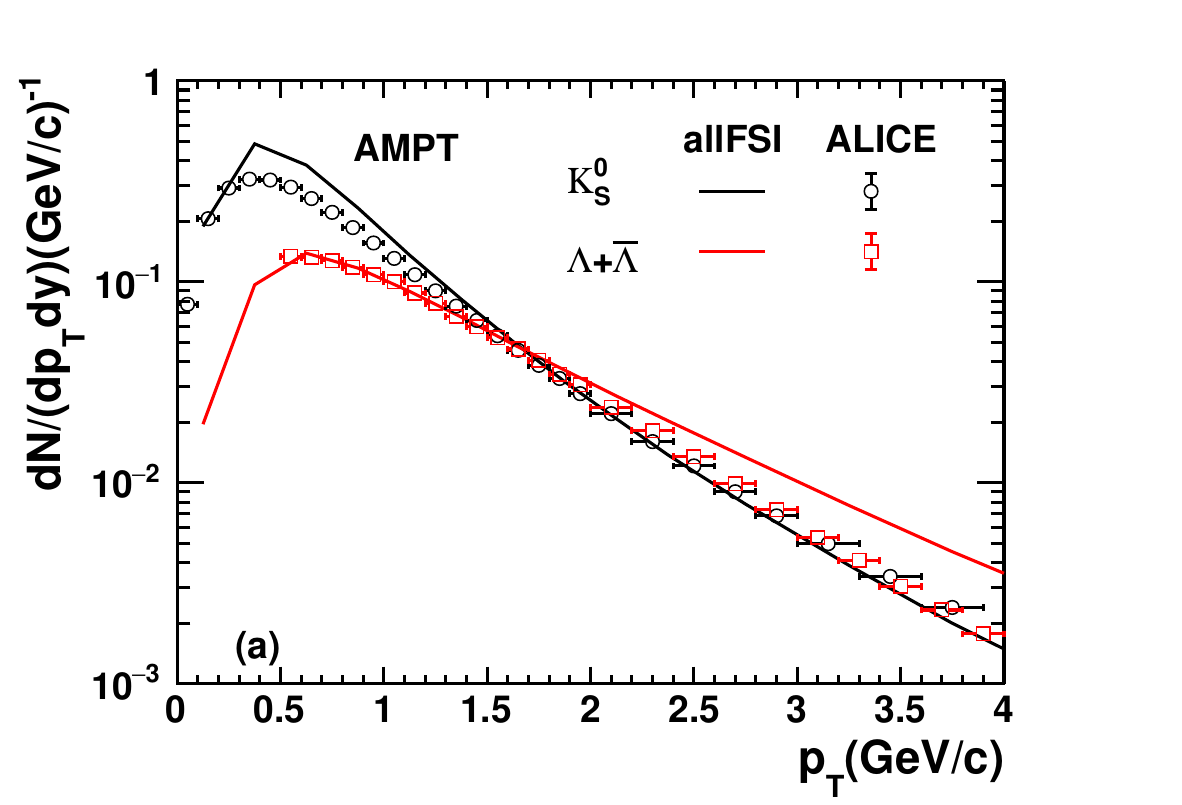}
\includegraphics[width=0.48\textwidth]{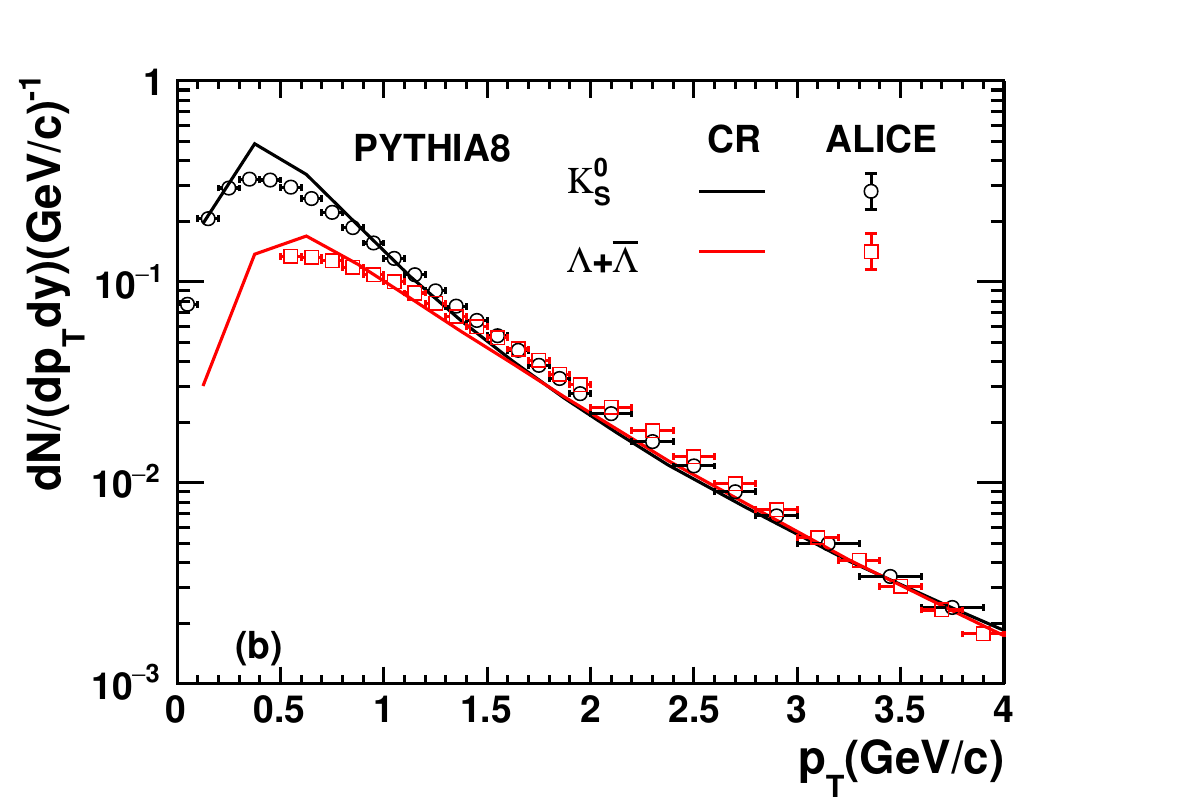}
    \caption{Transverse momentum spectra of $K^{0}_{S}$ (black) and $\Lambda+\bar{\Lambda}$ (red) in $\sqrt{s}=13$ TeV pp collisions at mid-rapidity $|y|<0.5$. The lines represent model calculations from AMPT with all final state interactions (a) and PYTHIA8 model with color reconnection effects (b), while the open markers represent ALICE data~\cite{ALICE:2020jsh}.}
    \label{fig:pt_LKs}
\end{figure*}
 
 \begin{figure*}[hbt!]
    \centering
\includegraphics[width=0.48\textwidth]{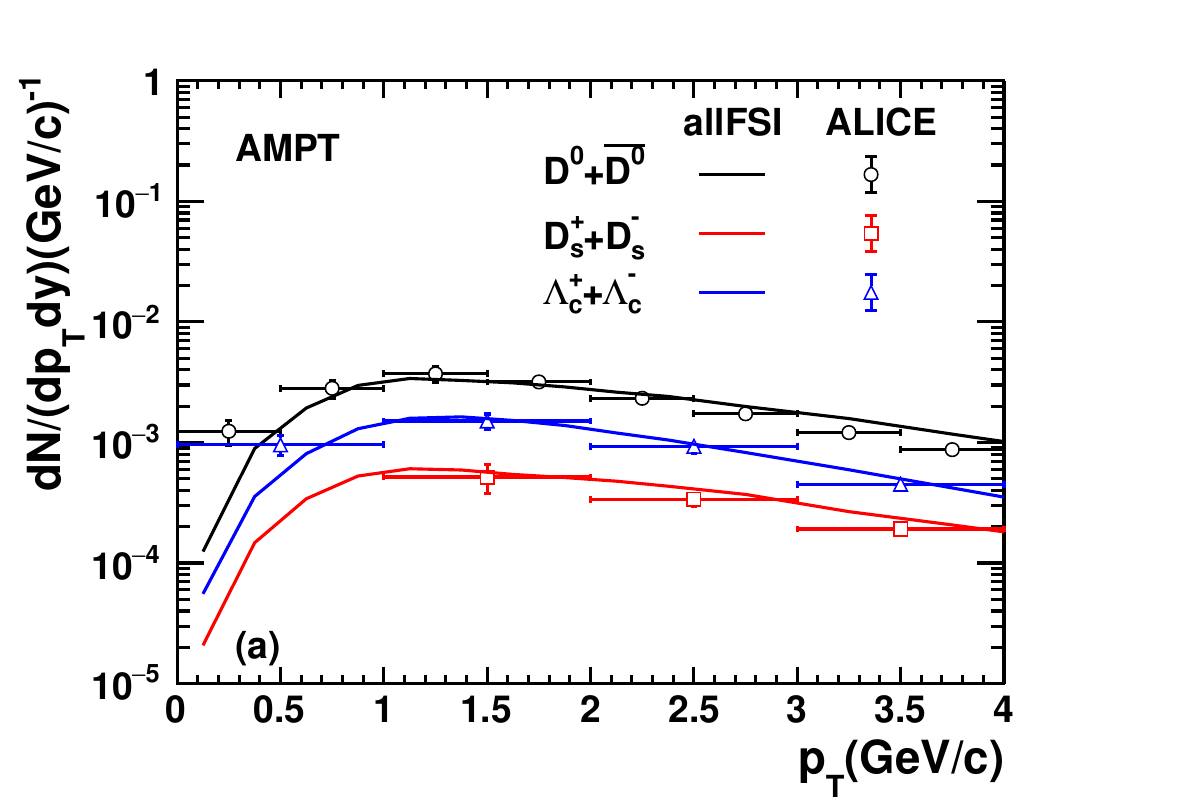}
\includegraphics[width=0.48\textwidth]{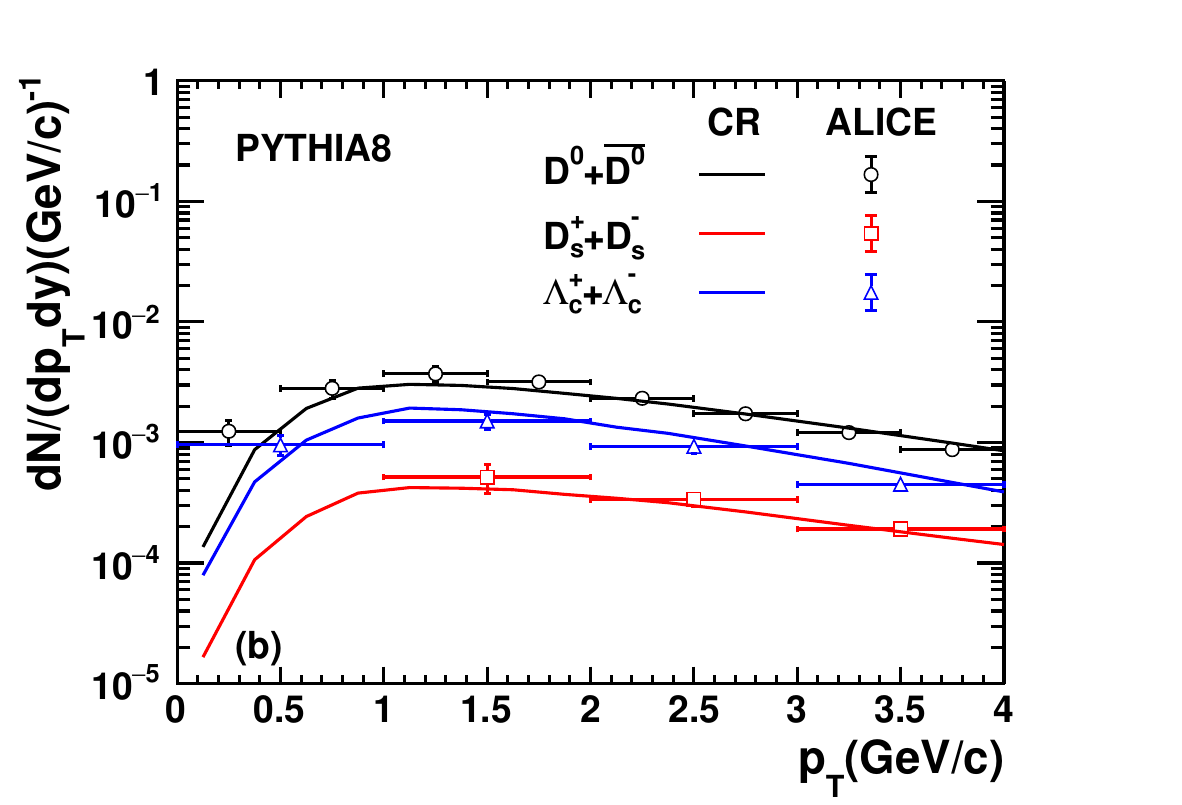}
    \caption{Transverse momentum spectra of $D^0+\bar{D^0}$ (black), $D_s^{\pm}$ (red), and $\Lambda_c^{\pm}$ (blue) in $\sqrt{s}=13$ TeV pp collisions at mid-rapidity $|y|<0.5$. The experimental data and model data in the figure are both the sum of particles and anti-particles divided by 2. The lines represent model calculations from AMPT with all final state interactions (a) and PYTHIA8 model with color reconnection effects (b), while the open markers represent ALICE data~\cite{ALICE:2023sgl}.}
    \label{fig:pt_D0DsLc}
\end{figure*}

 \begin{figure*}[hbt!]
    \centering
\includegraphics[width=0.48\textwidth]{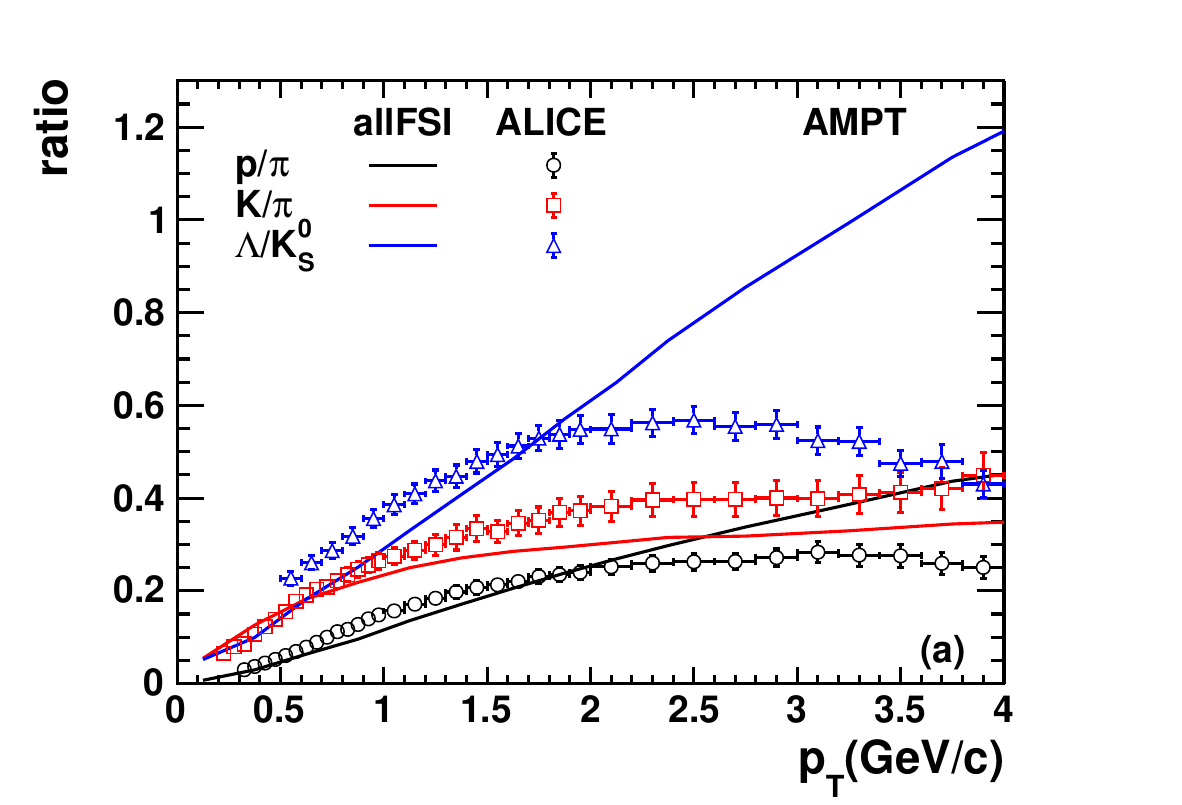}
\includegraphics[width=0.48\textwidth]{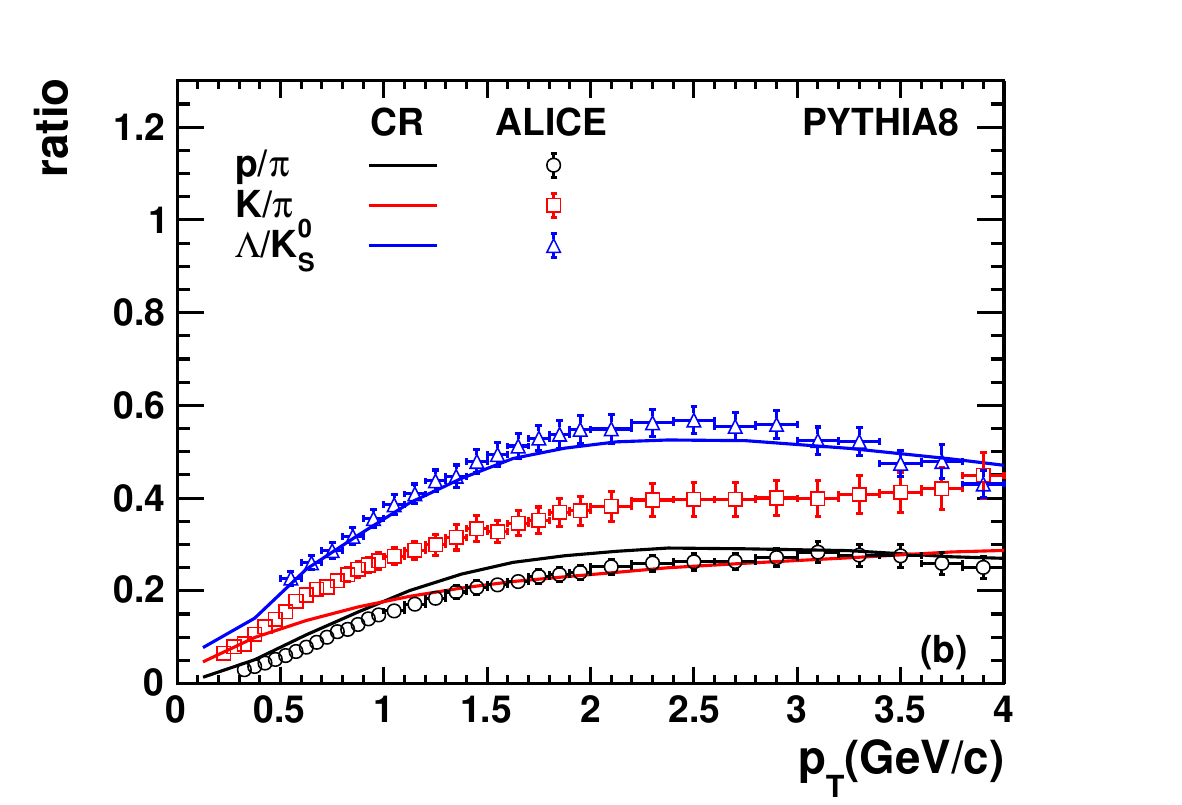}
    \caption{$p_{T}$ -differential p/$\pi$ ratio (black), K/$\pi$ ratio (red), and $\Lambda$/$K_S^0$ ratio (blue) in $\sqrt{s}=13$ TeV pp collisions at mid-rapidity $|y|<0.5$. Here, $\Lambda/K_S^0$ is actually $(\Lambda + \bar{\Lambda}) / 2K_S^0$. The lines represent model calculations from AMPT with all final state interactions (a) and PYTHIA8 model with color reconnection effects (b), while the open markers represent ALICE data~\cite{ALICE:2020jsh}.}
    \label{fig:pt_ratio_light}
\end{figure*}

  \begin{figure*}[hbt!]
    \centering
\includegraphics[width=0.48\textwidth]{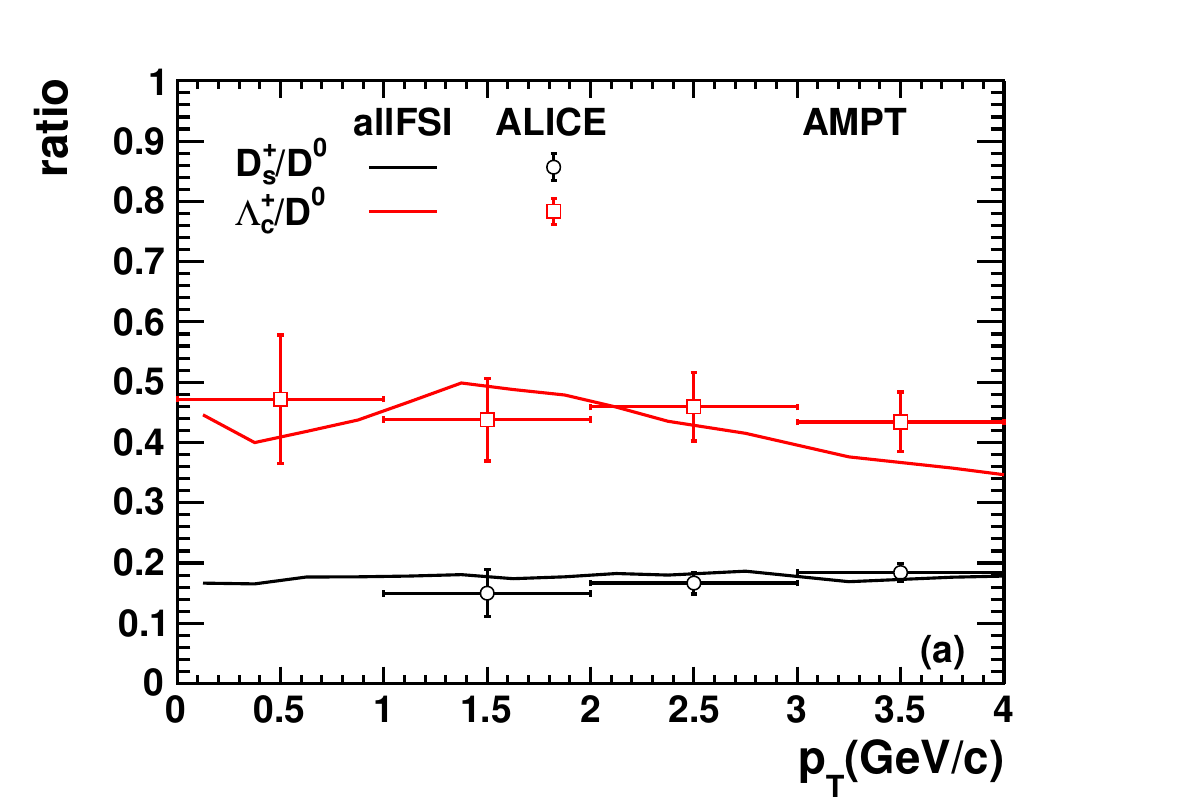}
\includegraphics[width=0.48\textwidth]{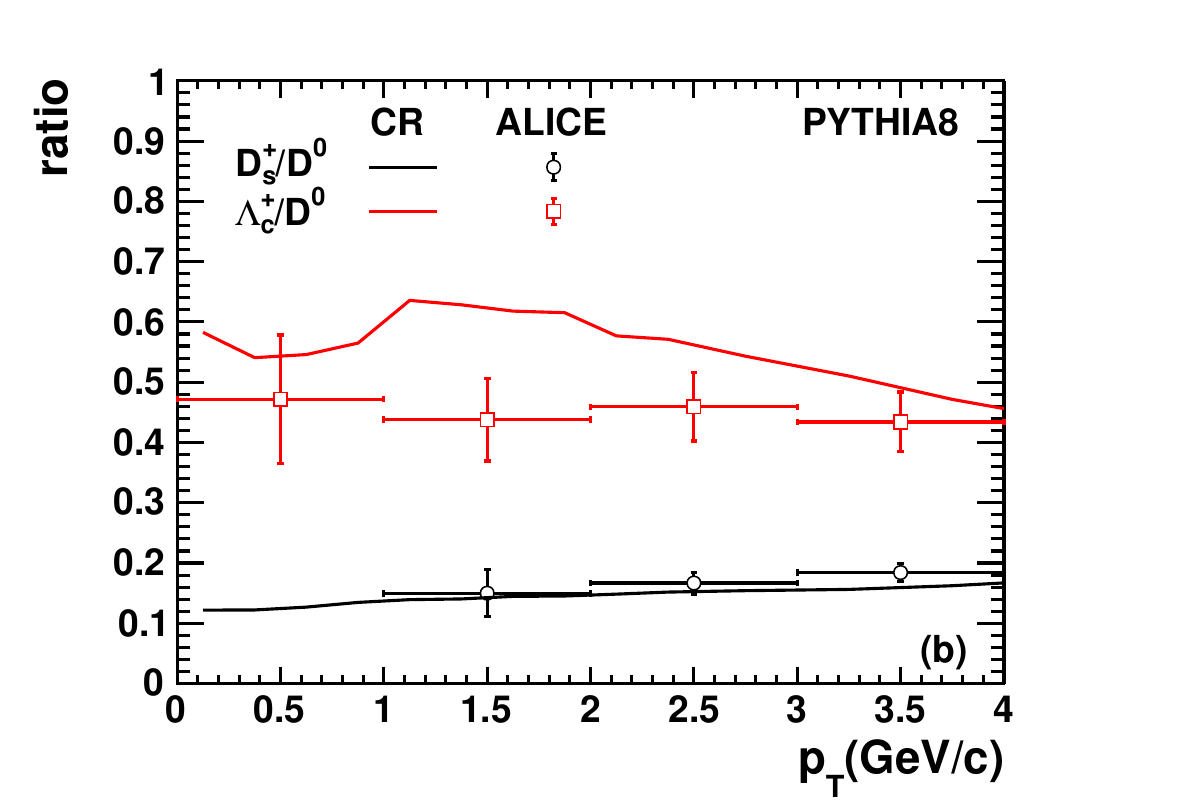}
    \caption{$p_{T}$ -differential $D_s^+$/$D^0$ ratio (black) and $\Lambda_c^+$/$D^0$ ratio (red) in $\sqrt{s}=13$ TeV pp collisions at mid-rapidity $|y|<0.5$. The lines represent model calculations from AMPT with all final state interactions (a) and PYTHIA8 model with color reconnection effects (b), while the open markers represent ALICE data~\cite{ALICE:2023sgl}.}
    \label{fig:pt_ratio_heavy}
\end{figure*}

 \section{\label{sec:results}Results}
	
This work is focusing on the hadron productions in pp collisions with $\sqrt{s} = 13$ TeV at midrapidity ($|y| < 0.5$) for eight different particle species: $\pi^{\pm}$, $K^{\pm}$, p($\bar{p}$), $\Lambda(\bar{\Lambda})$, $K_S^0$, $D^0$, $D_s^{\pm}$, and $\Lambda_c^{\pm}$. In Sec.~\ref{subsec:inclusive} we present the inclusive productions and their ratios for these particle species in minbias events from AMPT and PYTHIA color reconnection model with comparisons to the ALICE experimental data~\cite{ALICE:2020jsh,ALICE:2023sgl}. The multiplicity dependent structures of these particle species ratios are further examined in Sec.~\ref{subsec:mult} and Sec.~\ref{subsec:dbratio}. The events are classified based on the number of particles accepted by the ALICE V0 detector acceptance within the pseudo-rapidity regions $-3.7 < \eta < -1.7$ and $2.8 < \eta < 5.1$ following the prescription in Refs.~\cite{ALICE:2020nkc,ALICE:2019avo}. The production of these identified particles has been studied with respect to the variation of the charged particle density in each event class also using the same percentile definition. To illustrate the final state interaction effects embedded in the AMPT model, we present the comparisons of the AMPT results with parton and hadron transport stage switched on sequentially labeled as noFSI, pFSI and allFSI.

	\subsection{Inclusive particle production}\label{subsec:inclusive}

        The data for all light-flavor particles in this paper represent the sum of particles and anti-particles, while the data for heavy-flavor particles are the sum of particles and anti-particles divided by 2, following experimental analysis conventions. For the inclusive hadron productions, the events are selected according to the minbias trigger used in the ALICE experiment requiring signals accepted by either side of the V0 detector. The transverse momentum spectra for charged pion, kaon and proton with $|y|<0.5$ are shown in Fig.~\ref{fig:pt_pikp} produced in inelastic pp collisions at $\sqrt{s}=13$ TeV. It is shown that the ALICE data of pion and kaon spectra can be roughly described by the allFSI model with a slight overestimation of pion productions at $p_T$ around 1 GeV$/c$. On the other hand, the proton spectra is overestimated by allFSI model with $p_T$ above 1.5 GeV$/c$. The overestimation to proton is because the current coalescence model in AMPT can affect the hadrons even in the high $p_T$ region. The coalescence overpredicts baryon spectra at high $p_T$ because it assumes quarks combine directly by adding their momenta, neglecting the dominance of fragmentation at high $p_T$, where baryon production becomes less efficient. In addition, the hadronic rescattering, which brings some radial flow like effect to proton spectra also enhances the proton production at higher $p_T$. More similar comparisons will be shown in the Sec.~\ref{subsec:dbratio}. The CR model results can describe the kaon $p_T$ distribution reasonably while the pion spectra are overpredicted even to the higher $p_T$ region. The proton yield is significantly overestimated on the whole $p_T$ regime due to the enhanced junction formations in the BLC CR model. This enhanced junction formation can lead to significantly enlarged baryon productions with different flavor components and need to be regulated with the diquark production parameter in the string fragmentation model as discussed in Ref.~\cite{Bierlich:2023okq}. 

        We compare the transverse momentum dependent of neutral strange hadron $K_S^0$ and $\Lambda$ productions in Fig.~\ref{fig:pt_LKs}. The strange meson $K_S^0$ spectra are well described in both allFSI and color reconnection model with slight overestimation around $p_T$ of 0.5 to 1 GeV$/c$ at a similar level. The $\Lambda$ $p_T$ distribution has been well described by the color reconnection model, while allFSI overestimates the yield at higher $p_T$. The overestimation in allFSI is largely connected to the application of the coalescence mechanism over the entire transverse momentum region. The overall yield of $\Lambda$ in allFSI roughly agrees with the experimental data due to the sensible choice of $r_{BM}^{s}$ in strange baryon production.

        The transverse momentum spectra for $D^0$, $D_s^{\pm}$, and $\Lambda_c^{\pm}$ are shown in Fig.~\ref{fig:pt_D0DsLc}. We convert the ALICE data from $d^2\sigma/dp_Tdy$ to $d^2N/dp_Tdy$ using the inelastic cross section $\sigma_{inel}=77.6$ mb estimated in~\cite{Loizides:2017ack} to make a consistent comparison with our model calculations. It is shown that the $p_{T}$ distributions for all three charmed hadron species are well described in the allFSI model with refined $r_{BM}$ parameter for charm baryon formation. Both the calculations from the allFSI model and the color reconnection model agree with the experimental data in a reasonable way.  

        The particle ratios $p/\pi$, $K/\pi$ and $\Lambda/K_S^0$ varying with the transverse momentum are presented in Fig.~\ref{fig:pt_ratio_light}. The baryon to meson ratios in allFSI for both $p/\pi$ and $\Lambda/K_S^0$ are consistent with the experimental data within $p_T$ range less than 2 GeV$/c$ but significantly overestimated at higher $p_T$. This observation is an outcome of our implementation for the quark coalescence hadronization mechanism over the entire $p_T$ range. The strange to non-strange meson ratio $K/\pi$ is slightly underestimated at $p_T$ greater than 1 GeV$/c$. Figure.~\ref{fig:pt_ratio_light}(b) shows that the color reconnection model provides a satisfactory description to the baryon to meson ratios for $p/\pi$ and $\Lambda/K_S^0$ over the examined $p_T$ region, emphasizing the importance of beyond leading color junction formation to the understanding of baryon productions in string fragmentation models. However, the $K/\pi$ ratio is significantly lower than data for $p_T$ greater than 0.5 GeV$/c$ with CR effect.

        In Fig.~\ref{fig:pt_ratio_heavy}, the $p_T$ dependence of heavy-flavored particle ratios, namely $D_s^+/D^0$ and $\Lambda_c^+/D^0$, are shown. For the $D_s^+/D^0$ ratio, both allFSI results and the color reconnection model results agree with the ALICE data within uncertainties. No sizable $p_T$ dependence of the $D_s^+/D^0$ ratio can be found in both models within the examined $p_T$ range. For the charmed baryon to meson ratio $\Lambda_c^+/D^0$, the allFSI model describes the experimentally observed magnitude within uncertainties while the color reconnection model predicts a slightly higher ratio. A characteristic peak appears around $p_T$ of 1.5 GeV$/c$ in the $\Lambda_c^+/D^0$ ratio from both the AMPT model with all final state interactions and the string fragmentation model with color reconnection effects, indicating the qualitative similarities of these two approaches.

       \subsection{Multiplicity dependence of particle ratio}\label{subsec:mult}

   \begin{figure*}[hbt!]
    \centering
\includegraphics[width=0.48\textwidth]{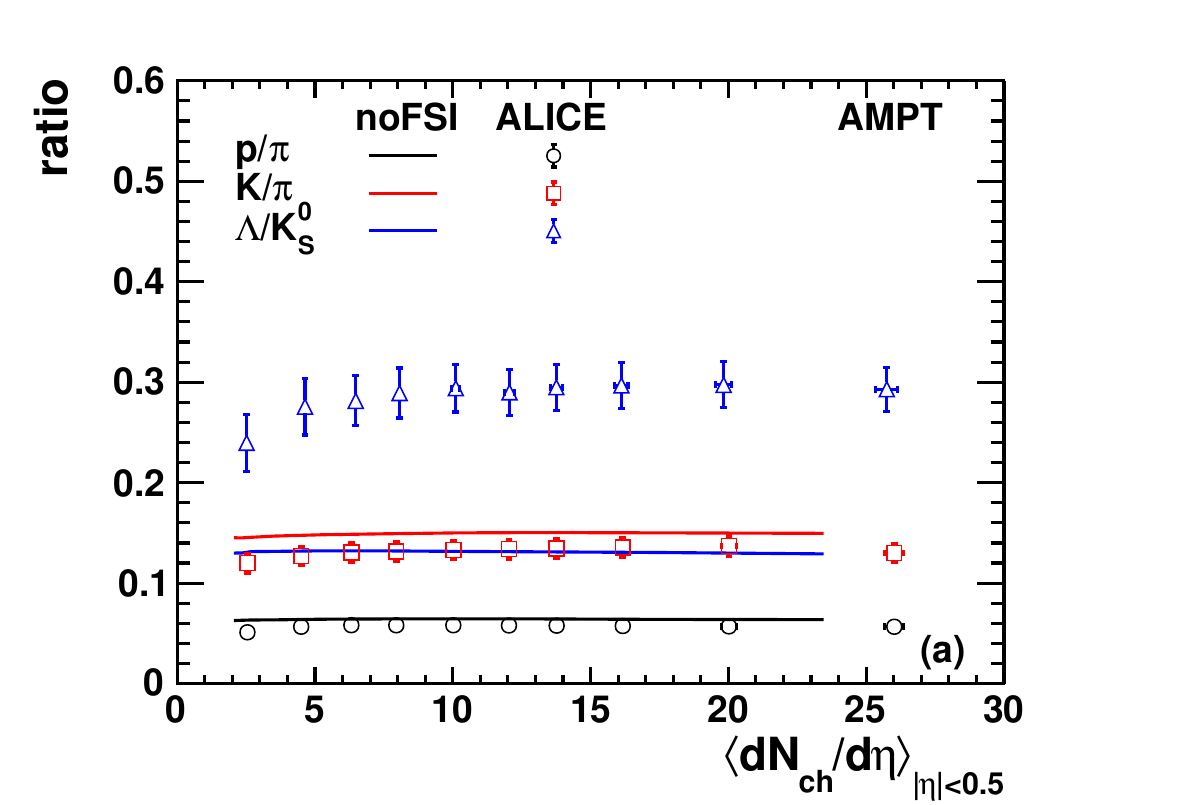}
\includegraphics[width=0.48\textwidth]{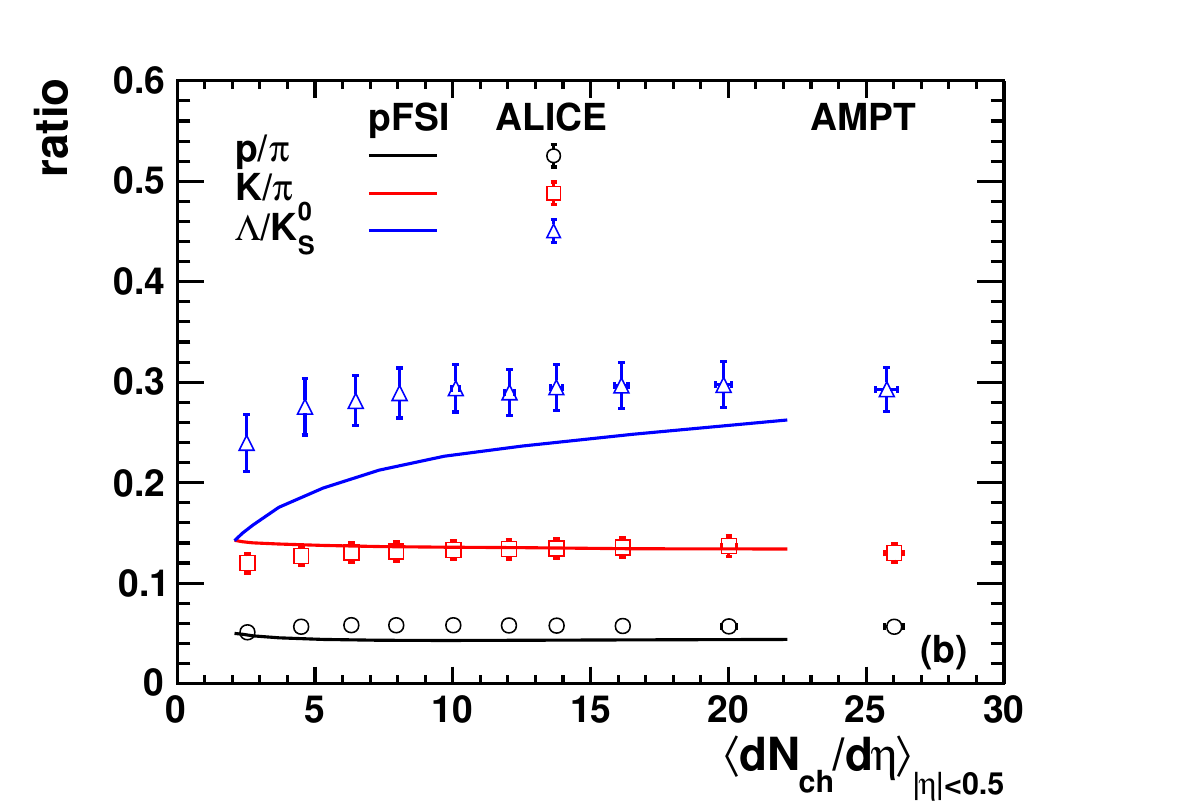}
\includegraphics[width=0.48\textwidth]{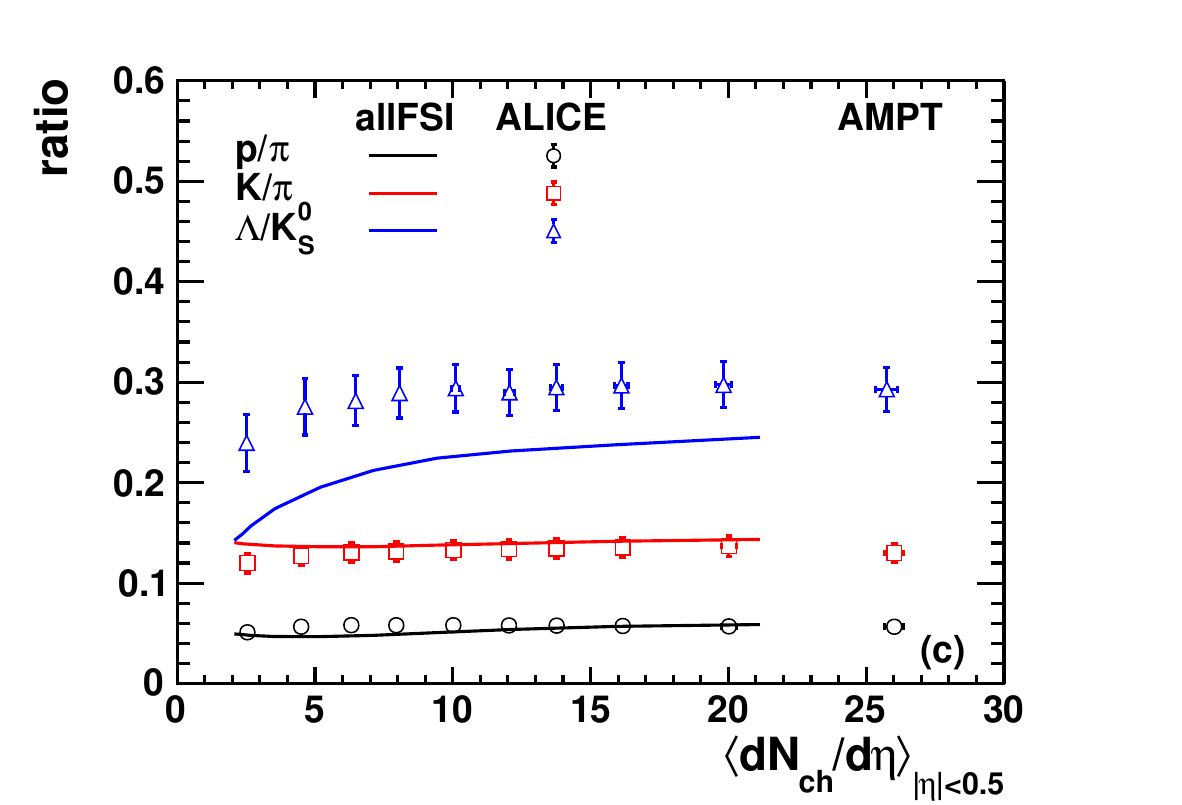}
\includegraphics[width=0.48\textwidth]{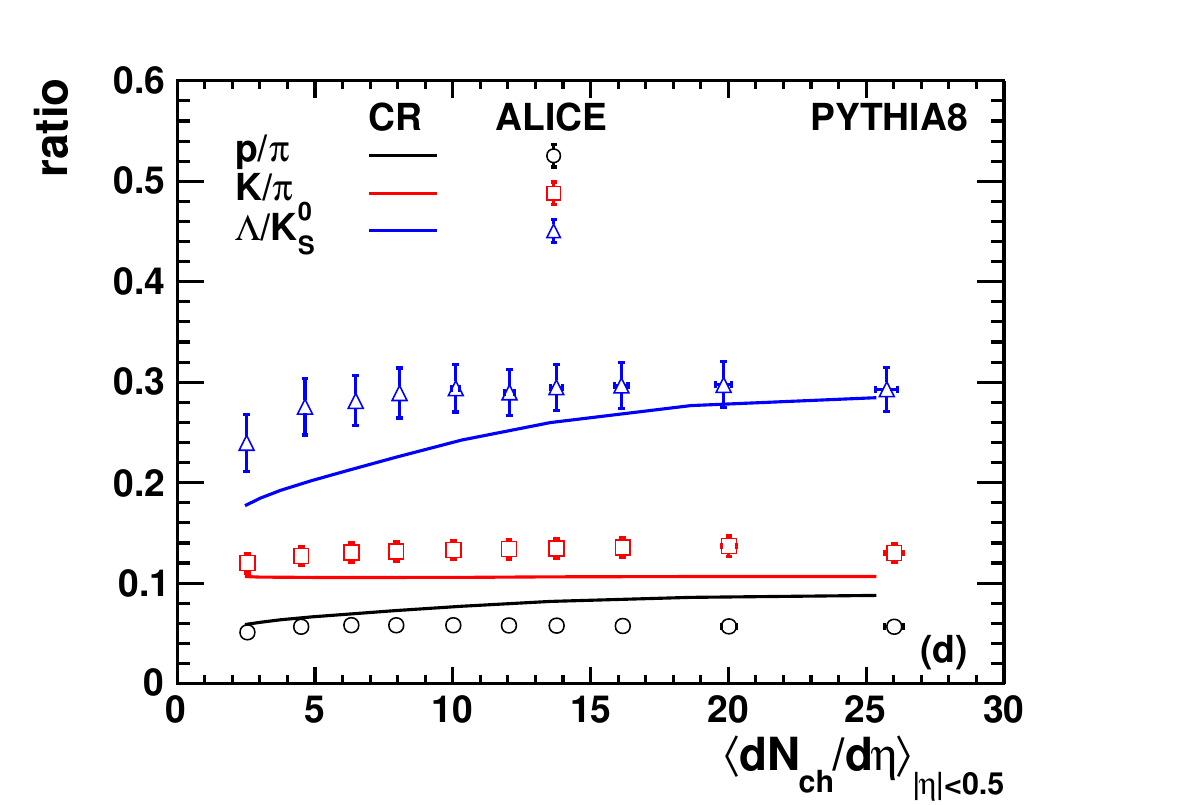}
    \caption{Integrated yield ratios of $K/\pi$ (red), $p/\pi$(black), and $\Lambda/K_S^0$ (blue) as a function of charged-particle multiplicity density in $\sqrt{s}=13$ TeV pp collisions at mid-rapidity $|y|<0.5$. Here, $\Lambda/K_S^0$ represents $(\Lambda + \bar{\Lambda}) / 2K_S^0$. The lines represent model calculations from AMPT noFSI(a), AMPT pFSI(b), AMPT allFSI(c), and PYTHIA8 CR(d), while the point markers represent ALICE data~\cite{ALICE:2020nkc,ALICE:2019avo}.}
    \label{fig:integrated light}
\end{figure*}      

   \begin{figure*}[hbt!]
    \centering
\includegraphics[width=0.48\textwidth]{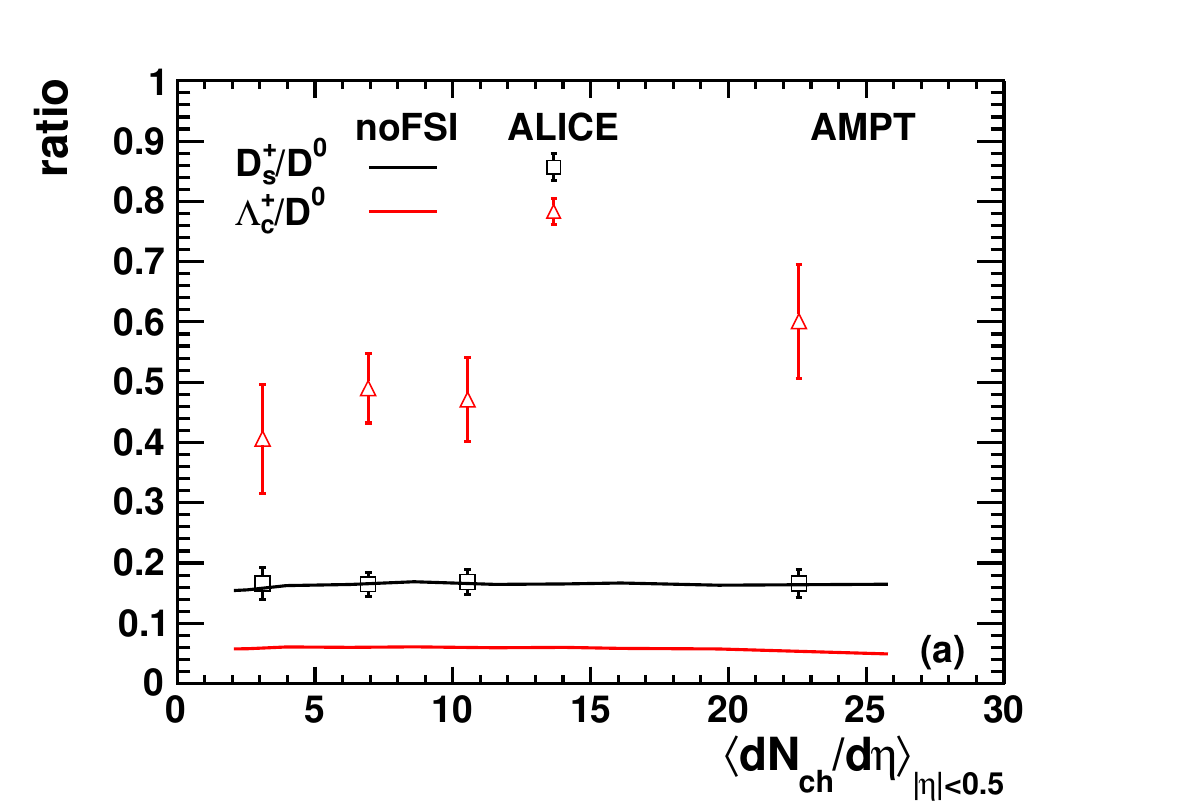}
\includegraphics[width=0.48\textwidth]{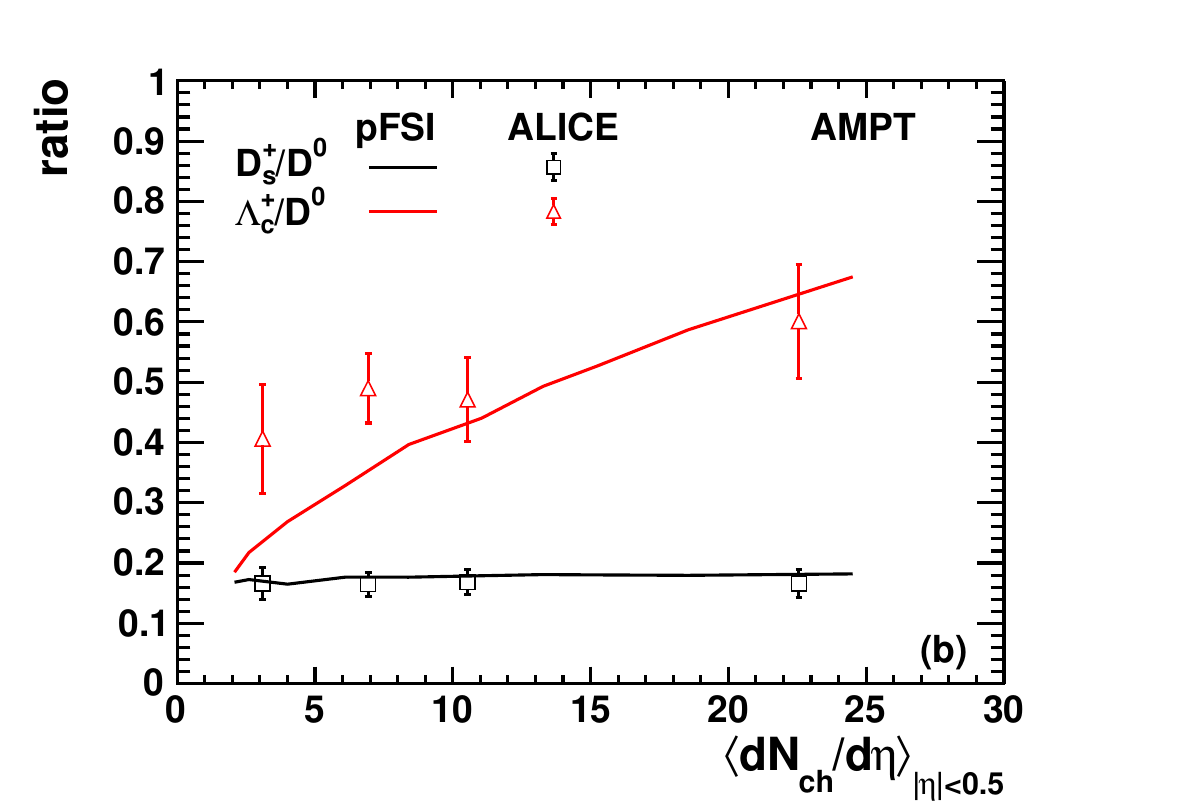}
\includegraphics[width=0.48\textwidth]{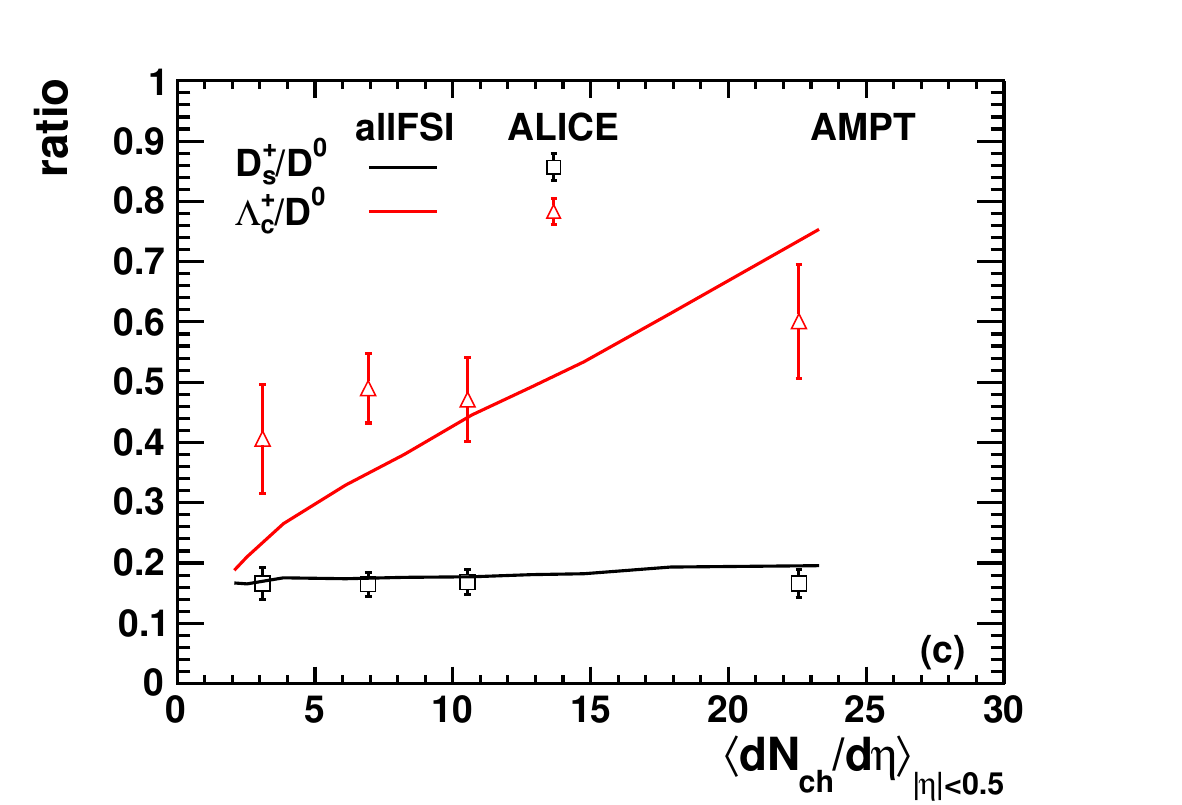}
\includegraphics[width=0.48\textwidth]{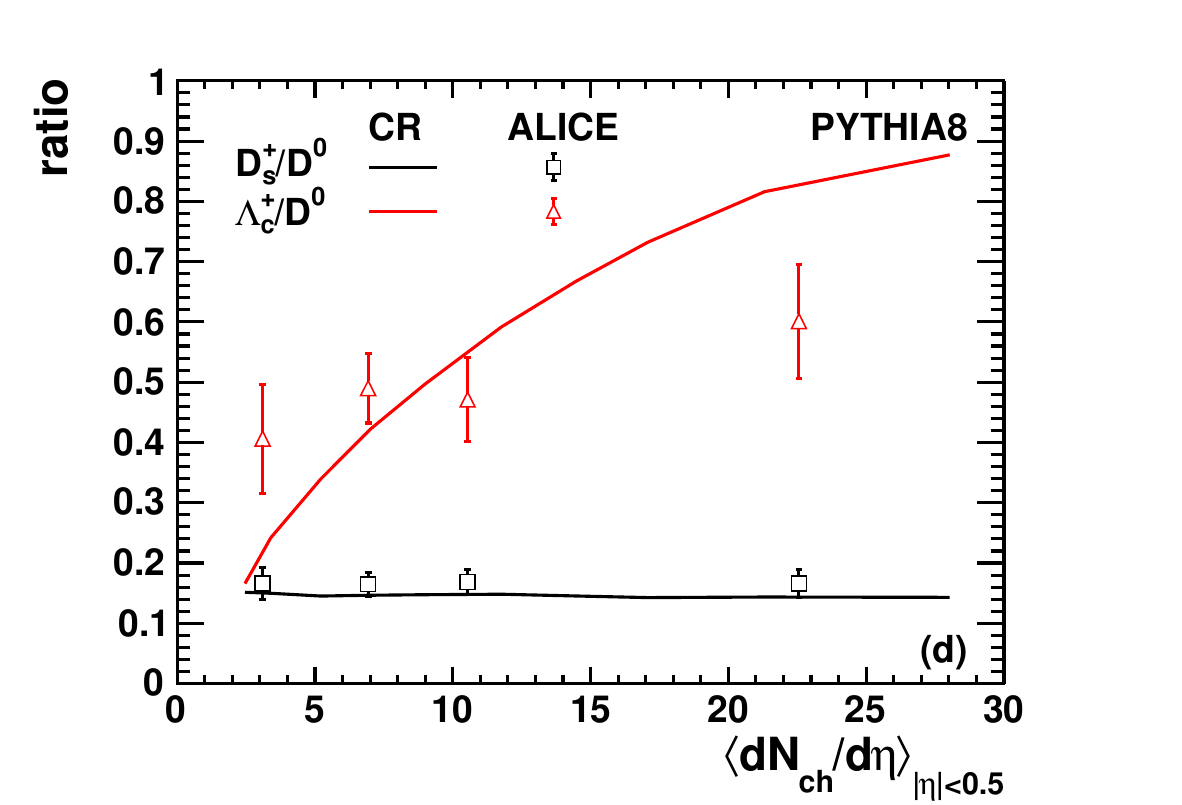}
    \caption{Integrated yield ratios of $D_s^+$/$D^0$ (black) and $\Lambda_c^+/D^0$ (red) as a function of charged-particle multiplicity density in $\sqrt{s}=13$ TeV pp collisions at mid-rapidity $|y|<0.5$. The lines represent model calculations from  AMPT noFSI(a), AMPT pFSI(b), AMPT allFSI(c), and PYTHIA8 CR(d), while the point markers represent ALICE data~\cite{ALICE:2021npz}.}
    \label{fig:integrated heavy}
\end{figure*}

   \begin{figure*}[hbt!]
    \centering
\includegraphics[width=0.48\textwidth]{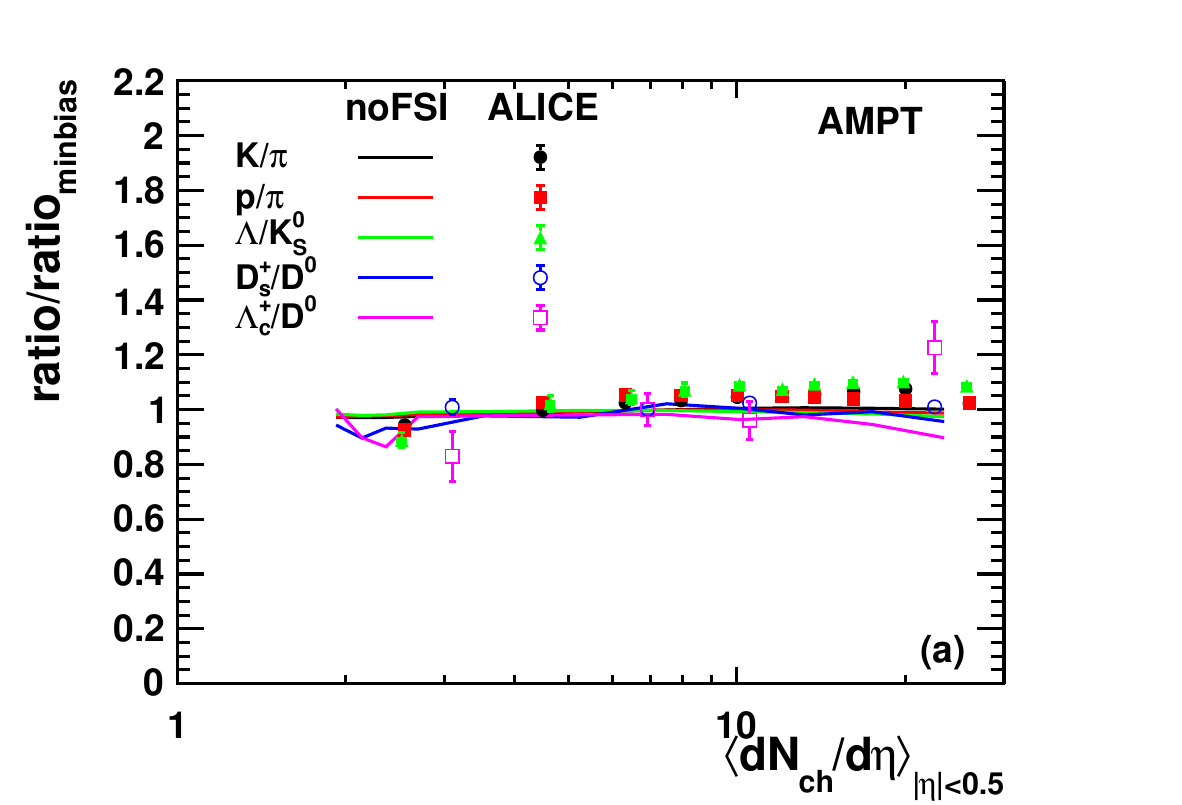}
\includegraphics[width=0.48\textwidth]{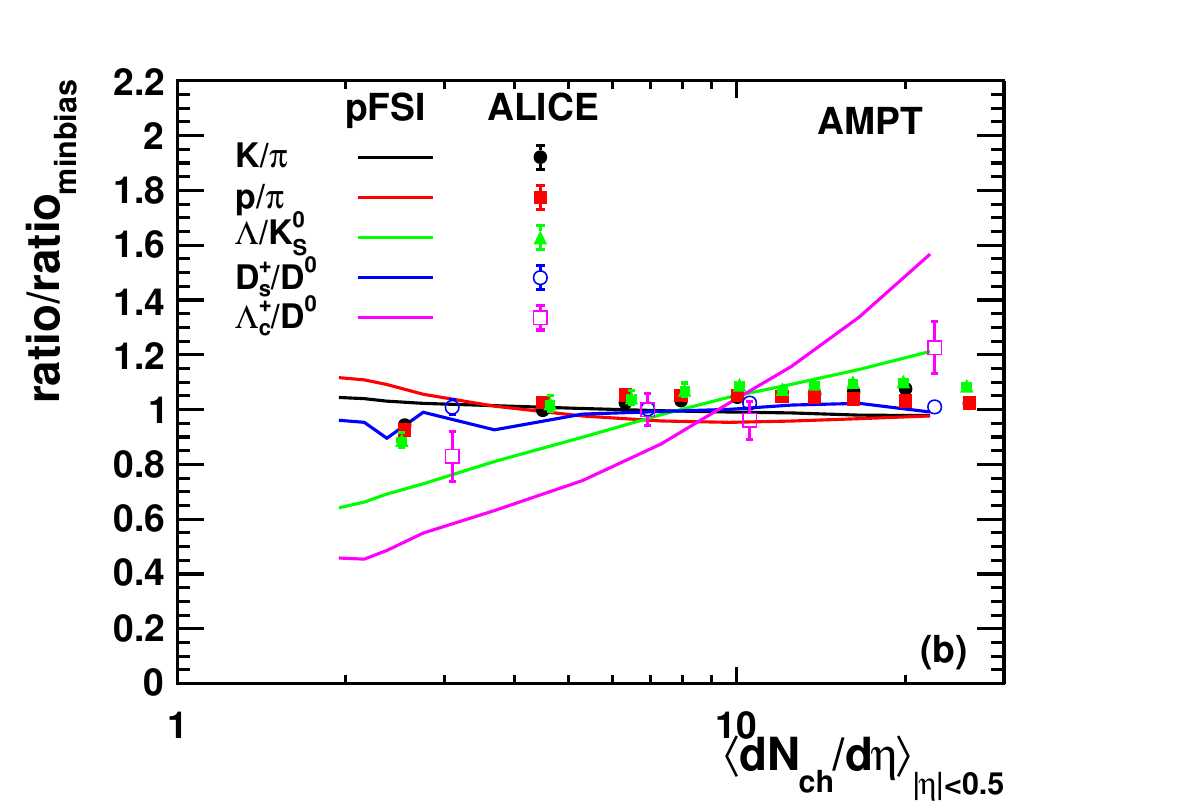}
\includegraphics[width=0.48\textwidth]{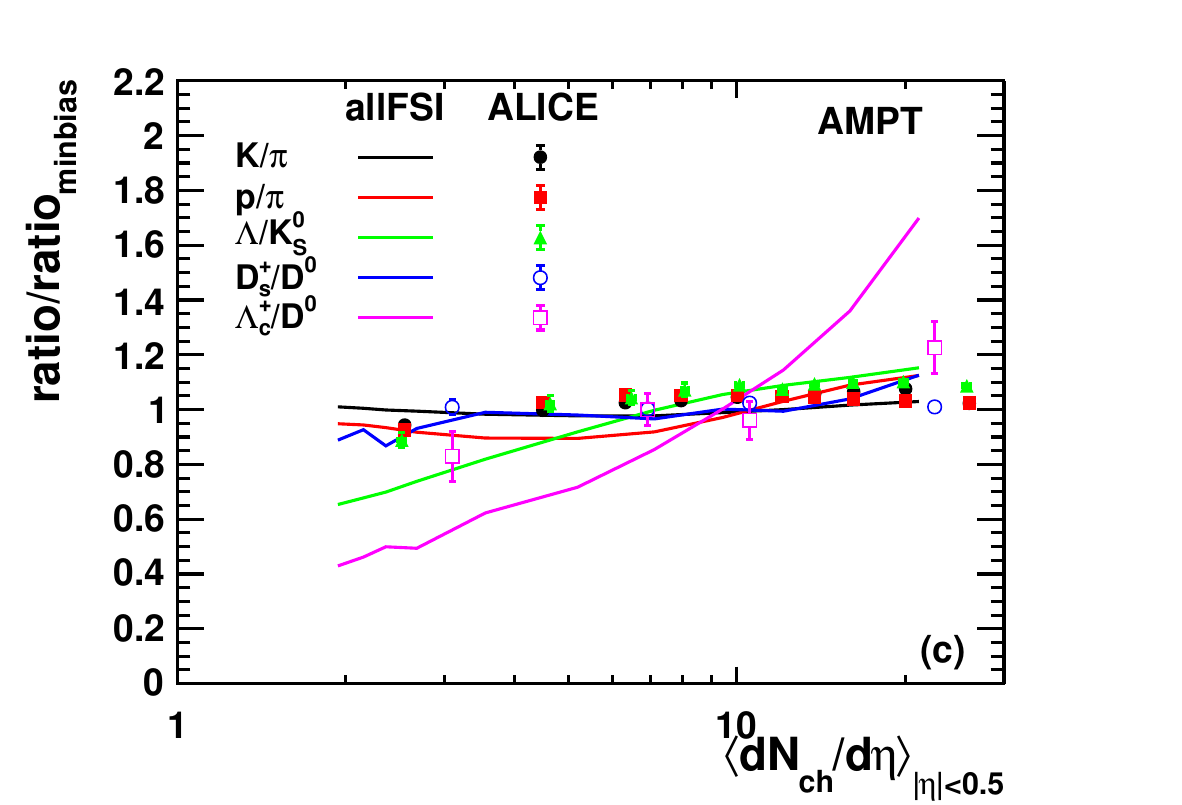}
\includegraphics[width=0.48\textwidth]{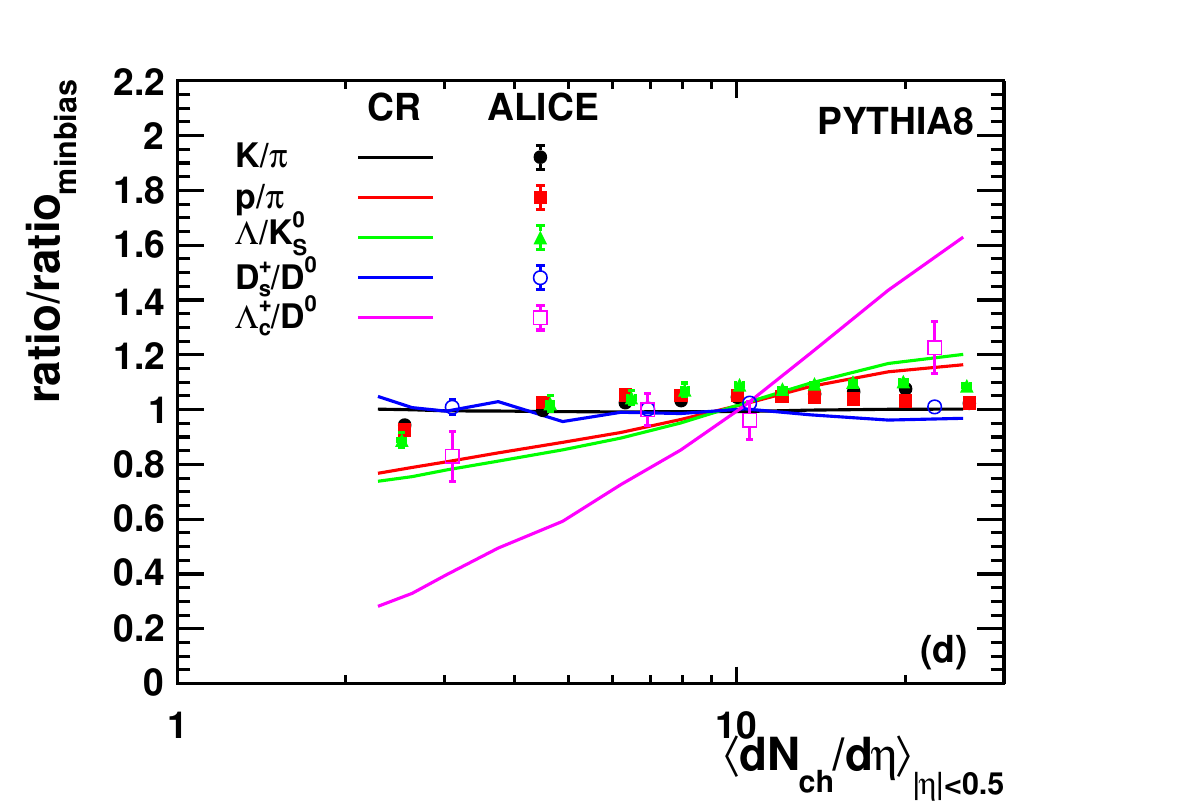}
    \caption{Integrated yield ratios normalized to the minbias particle ratio of p/$\pi$ (red), K/$\pi$ (black), $\Lambda$/$K_S^0$ (green), $D_s^+/D^0$ (blue) and $\Lambda_c^+/D^0$ (magenta) as a function of the charged particle density in $\sqrt{s}=13$ TeV pp collisions at mid-rapidity $|y|<0.5$. Here, $\Lambda/K_S^0$ represents $(\Lambda + \bar{\Lambda}) / 2K_S^0$. The lines represent model calculations from AMPT noFSI(a), AMPT pFSI(b), AMPT allFSI(c), and PYTHIA8 CR(d), while the markers represent ALICE data~\cite{ALICE:2021npz,ALICE:2020nkc,ALICE:2019avo}.}
    \label{fig:slope}
\end{figure*}   

        After examining the inclusive hadron productions, we will investigate the variations of the hadron chemical compositions by exploring the ratios of different particle species with the final state charged particle density in this section.
        Fig.~\ref{fig:integrated light} presents the $p_T$ integrated particle ratios of $K/\pi$, $p/\pi$, and $\Lambda/K_S^0$ as a function of $\langle dN_{ch}/d\eta \rangle$. To quantify the impact of different transport stages to the multiplicity dependence of hadron productions, we turn on the final state parton and hadron interactions in the AMPT model in a step-by-step way. It is shown in Fig.~\ref{fig:integrated light}(a) that no multiplicity dependence of the light flavor particle ratios will be generated if all final state interactions are turned off in the AMPT model. The $\Lambda/K_S^0$ ratio is found to grow rapidly with the event multiplicity at low $\langle dN_{ch}/d\eta \rangle$ and becomes almost saturated in the high $\langle dN_{ch}/d\eta \rangle$ region when final state parton interactions and the coalescence hadronization are included as shown in Fig.~\ref{fig:integrated light}(b). As indicated in Fig.~\ref{fig:integrated light}(c), further hadronic rescatterings slightly suppress the increase of $\Lambda/K_S^0$ ratio and lead to a weak $p/\pi$ multiplicity dependence. The AMPT model, including all final state interactions, is found to effectively reproduce the  multiplicity dependence of the light flavor particle ratios observed in the experimental data~\cite{ALICE:2020nkc,ALICE:2019avo}, except for underestimating the $\Lambda/K_S^0$ ratio. This underestimation can be understood as the initial conditions from PYTHIA8 provide too small strange baryon yield, which has also been shown in Fig.~\ref{fig:integrated light}(a). The final state interactions together with the coalescence hadronization mechanism are needed to describe the multiplicity dependence of light hadron productions in experiment especially for the strange baryon to meson ratio within the AMPT framework. On the other hand, the string fragmentation model with CR effects predicts a strong multiplicity dependence of the baryon to meson ratio both for $p/\pi$ and $\Lambda/K_S^0$ shown in Fig.~\ref{fig:integrated light}(d). The multiplicity dependence of the strange baryon to meson ratio found in experiment is well captured by the CR model while the $p/\pi$ ratio is overestimated and the $K/\pi$ ratio is lower than the experimental data.

        In Fig.~\ref{fig:integrated heavy}, we compare the charm hadron ratios $D_s^+/D^0$ and $\Lambda_c^+/D^0$ as a function of $\langle dN_{ch}/d\eta \rangle$ from both AMPT and CR model, compared to the experimental data. It is shown in this comparison that the $D_s^+/D^0$ ratio barely changes with the final state interactions implemented in the current AMPT and the measured $D_s^+/D^0$ ratio is well described by the model calculations. The $\Lambda_c^+/D^0$ ratio, nevertheless, is flat over the entire multiplicity range and significantly lower than the ALICE data if no final state interactions are included in the AMPT as shown in Fig.~\ref{fig:integrated heavy}(a). The incorporation of the parton rescatterings in AMPT leads to a significant multiplicity dependence of the $\Lambda_c^+/D^0$ ratio. The charmed baryon to meson ratio drastically grows with $\langle dN_{ch}/d\eta \rangle$, seemingly faster than the experimental data. The increase of this fraction can be understood that more charm quarks are involved in the partonic scatterings in high multiplicity events and will be hadronized via the coalescence process with a higher baryon formation probability. The string fragmentation model with CR effects also predicts a very fast increase of $\Lambda_c^+/D^0$ ratio varying with the event multiplicity and a smooth $D_s^+/D^0$ ratio consistent with the experimental data. 

        It is interesting to see that the $\Lambda_c^+/D^0$ ratio in AMPT seems to keep increasing with $\langle dN_{ch}/d\eta \rangle$ even in the very high multiplicity events, unlike the case observed in Fig.~\ref{fig:integrated light} for strange baryon to meson ratio calculations. In the string melting AMPT framework, the hadron productions are not only dependent on the coalescence hadronization mechanism but also related to fraction of unmelt initial strings in each event. In high multiplicity events, the initial strings are more likely to be destructed by the surrounding parton matter and the coalescence hadronization effect becomes more dominant. The saturation of strange baryon to meson ratio at intermediate to high multiplicity indicates that the initial strings containing strange quark have been fully destroyed and the hadron production ratios are determined mostly by the coalescence parameter value in this scenario. The charm initial string objects are generated independently from the bulk medium and may not fully participate in its evolution. A substantial fraction of undestroyed charm strings remains, though it gradually decreases with increasing event multiplicity. This discrepancy might suggest a different thermalization degree for charm and strange quarks in small system. It has been extensively discussed that the canonical suppression in statistical hadronization model may play a similar role in the determination of the multiplicity dependence for both light and heavy flavor hadron productions~\cite{Vislavicius:2016rwi,Chen:2020drg,Dai:2024vjy}, suggesting a suppressed baryon to meson ratio at low multiplicity and a saturated value in high multiplicity events even for charm and bottom hadrons. Testing the multiplicity dependence of heavy flavor hadron production ratios with high precision experimental data will be important to understand the details of these flavor hierarchy structures.

        In order to further identify the multiplicity dependent shape of the particle ratios with different quark flavor components, we present the self-normalized particle ratios for $p/\pi$, K/$\pi$, $\Lambda$/$K_S^0$, $D_s^+/D^0$ and $\Lambda_c^+/D^0$ in each multiplicity event class divided by the minbias particle ratio as a function of the charged particle density $\langle dN_{ch}/d\eta \rangle$ in Fig.~\ref{fig:slope}. In this way, the difference in the magnitude of the particle ratios with different flavors will be cancelled and the shape of each particle ratio can be quantitatively compared on the same ground indicating how fast a certain particle ratio changes with event multiplicity. It is shown in Fig.~\ref{fig:slope}(a) that the self-normalized ratios for all particle species are consistent with unity and independent of the charged particle density if all final state parton and hadron interactions are switched off in AMPT. The multiplicity dependence of the hadron ratios appears after the partonic level evolution stage is included as shown in Fig.~\ref{fig:slope}(b). Within the coalescence procedure of the AMPT model, a clear flavor hierarchy structure is produced, indicating an increased slope from $p/\pi$, $\Lambda/K_S^0$ to $\Lambda_c^+/D^0$ ratio, which depends on the final state multiplicity. The strange to non-strange meson ratio is flat across the entire multiplicity range. The follow-up hadronic rescatterings shown in Fig.~\ref{fig:slope}(c) leads to a slight modification to the multiplicity dependence of the self-normalized ratio for $p/\pi$ and $\Lambda/K_S^0$, while the flavor hierarchy structure is unchanged. The experimental data are converted from the measured particle ratios in each multiplicity bin divided by the corresponding ratio obtained for the minbias events incorporating all the events with different multiplicity bins~\cite{ALICE:2021npz,ALICE:2020nkc,ALICE:2019avo}. It is shown in the converted experimental data that the slopes for all the particle ratios are quite small in contrast to the model calculations, although a tantalizing flavor hierarchy structure exists in the three baryon to meson ratios. We also examine the same self-normalized particle ratio results from the color reconnection model calculations in Fig.~\ref{fig:slope}(d). A much larger slope is also predicted for the baryon to meson ratios with different quark flavors by the color reconnection model compared to the experimental data. However, the baryon to meson ratios $\Lambda/K_S^0$ and $p/\pi$ are found to be quite similar in this framework. The ordering between the non-strange and strange baryon to meson ratio obtained with the color reconnection model is not so clear as that found with the coalescence hadronization process.

       \subsection{Transverse momentum dependence of the double ratio} \label{subsec:dbratio}

   \begin{figure*}[hbt!]
    \centering
\includegraphics[width=0.48\textwidth]{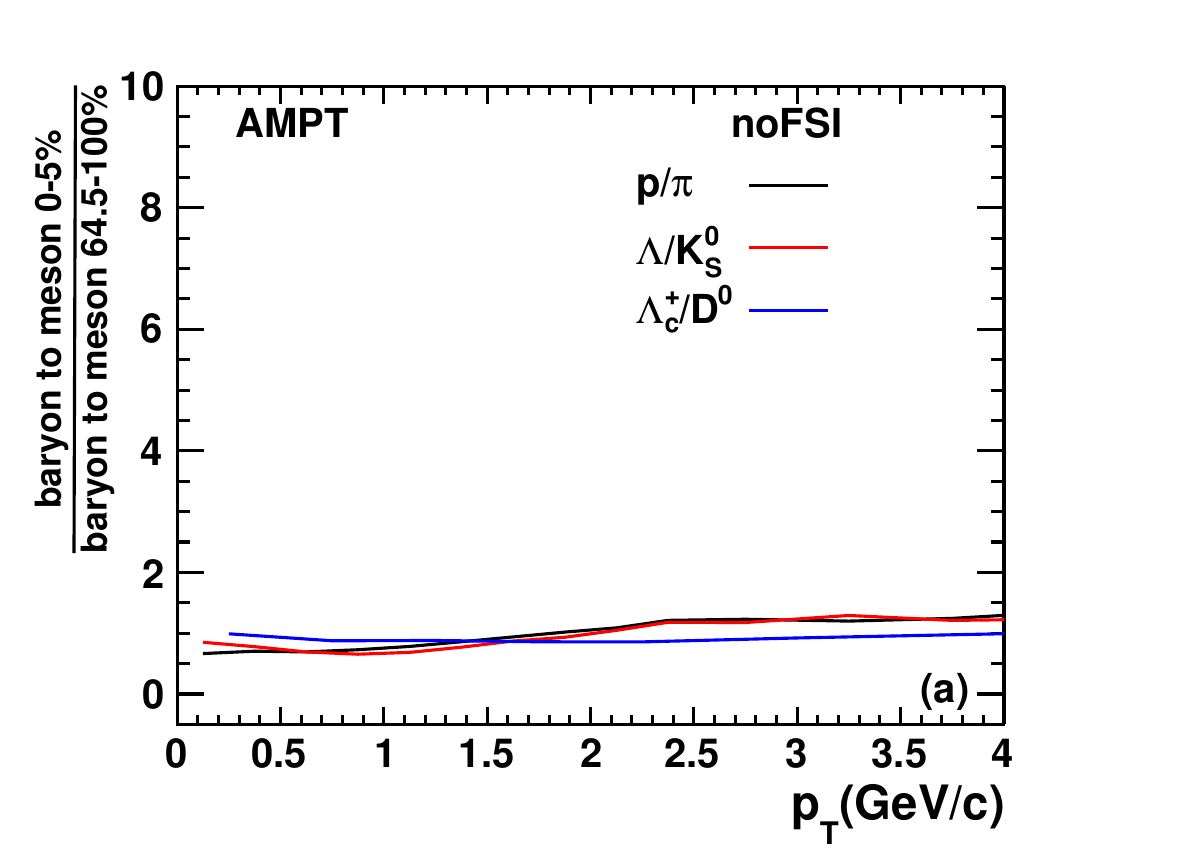}
\includegraphics[width=0.48\textwidth]{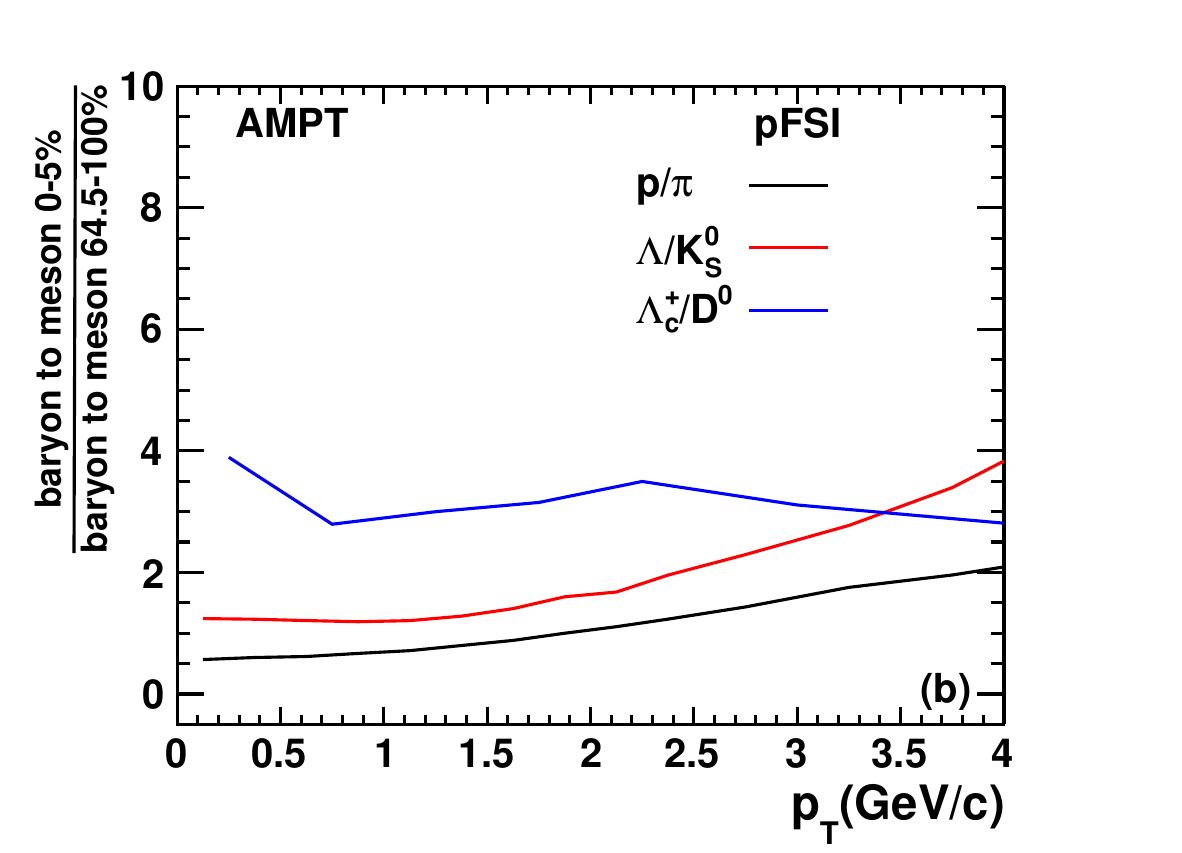}
\includegraphics[width=0.48\textwidth]{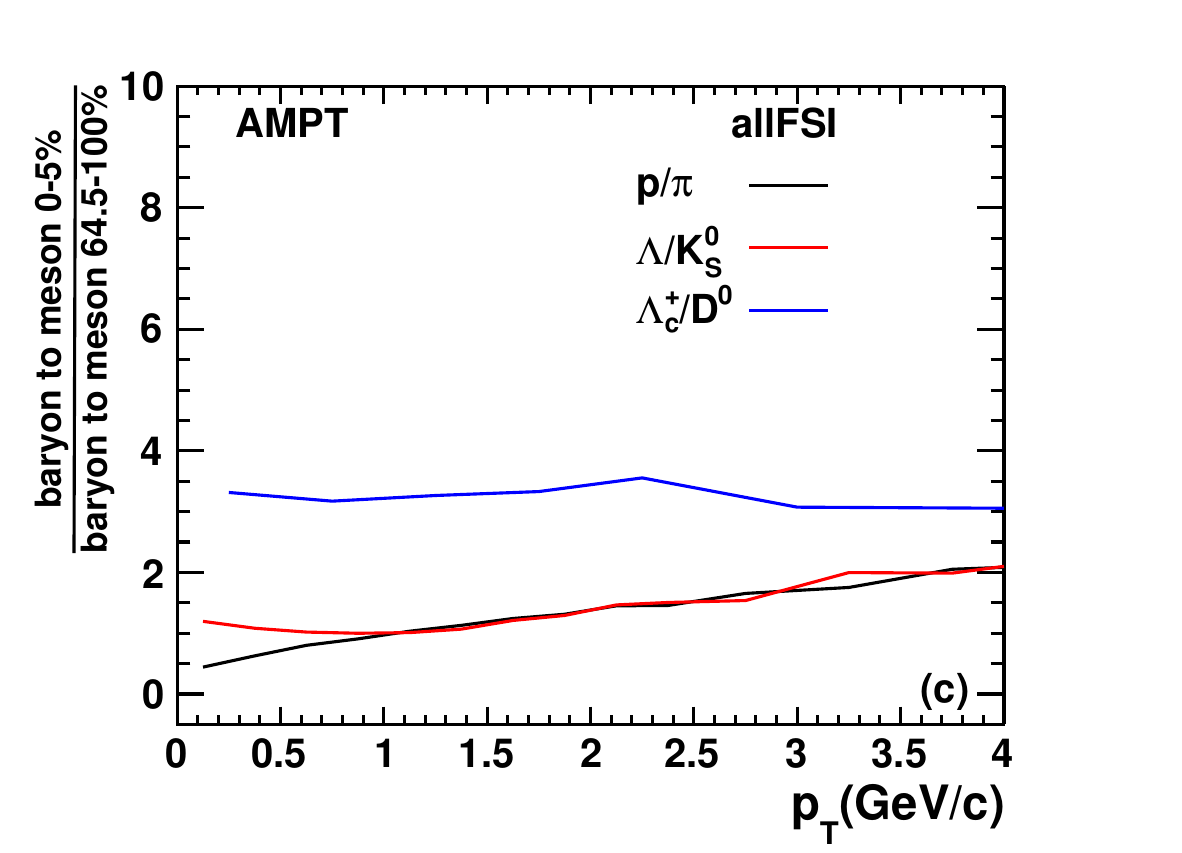}
\includegraphics[width=0.48\textwidth]{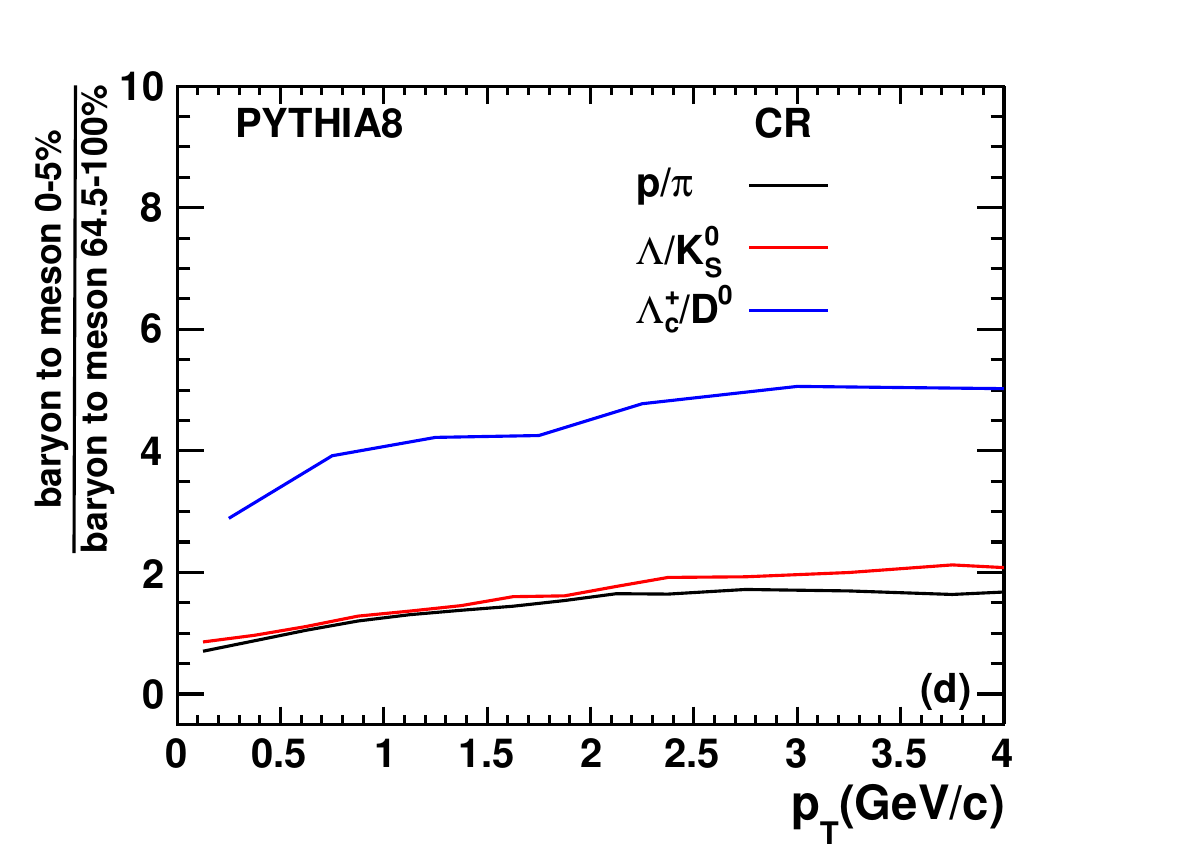}
    \caption{The $p_{T}$ dependence of the double ratios, defined as the particle ratios in high multiplicity events (0-5\%) divided by those in low multiplicity events (64.5-100\%), for p/$\pi$ (black), $\Lambda$/$K_S^0$ (red) and $\Lambda_c^+/D^0$ (blue) in $\sqrt{s}=13$ TeV pp collisions at mid-rapidity $|y|<0.5$. The lines represent model calculations from AMPT noFSI(a), AMPT pFSI(b), AMPT allFSI(c), and PYTHIA8 CR(d).}
    \label{fig:double}
\end{figure*}

        We further investigate the transverse momentum dependent modifications to the hadron production ratios induced by the event multiplicities and present the double ratios constructed with the particle ratios from central collisions (0-5\% centrality) divided by those from peripheral collisions (64.5-100\% centrality) in Fig.~\ref{fig:double}. As the strange to non-strange meson ratios $K/\pi$ and $D_s^+/D^0$ are found to be around unity, independent of the event multiplicity and $p_T$ in our model study, we only show the baryon to meson ratio results in this comparison. It is consistent with the expectations that all the particle ratios are close to unity when the final state parton and hadron interactions are off. The final state parton evolution and the corresponding coalescence hadronization process brings in non-trivial $p_T$ dependence for three baryon to meson ratios with different flavors and a clear ordering structure at low $p_T$ shown in Fig.~\ref{fig:double}(b). Within the investigated $p_T$ range, the $\Lambda/K_S^0$ and $p/\pi$ double ratios increase with $p_T$ while $\Lambda_c^+/D^0$ is almost unrelated to $p_T$. A significant ordering from $\Lambda_c^+/D^0$ to $p/\pi$ is observed especially at the lower $p_T$ region. The follow-up hadronic scattering suppresses the $\Lambda/K_S^0$ double ratio significantly at $p_T$ greater than 1 GeV$/c$ while the low $p_T$ region still preserves the flavor hierarchy structure shown in Fig.~\ref{fig:double}(c). It is also seen that the color reconnection model gives an increasing behavior dependent on $p_T$ for all baryon to meson ratios. The charm baryon to meson ratio is found to be enhanced much stronger than the light flavor sectors but no significant difference is observed between the $\Lambda/K_S^0$ and $p/\pi$ ratios. The charm baryon to meson ratio is more sensitive to the color reconnection effect in high multiplicity events compared to the light flavor sectors are largely related to the fact that heavy flavor baryons can be only produced with the baryon junction formations which are controlled by the reconnection probability of the string objects at beyond leading color level~\cite{Christiansen:2015yqa}. It is also interesting to see that there is some radial flow like feature in the $\Lambda_c^+/D^0$ ratio in high multiplicity pp collisions from color reconnection calculations especially in the low $p_T$ region, similar to the observations found in experimental data~\cite{ALICE:2021npz}. This behavior can be related to the junction formation process at beyond leading color level. As the charmed hadrons cannot be created in the fragmentation process, the enhanced $\Lambda_c^+$ production in color reconnection is always related to the junctions connecting two dipole strings. In low multiplicity events, the dipole strings are usually parallel to the beam direction, and the CR induced $\Lambda_c^+$ therefore favors the production in the low $p_T$ region~\cite{Bierlich:2023okq}. On the other hand, more high $p_T$ jets are produced in the high multiplicity events, leading to the reconnected strings shifted to the higher $p_T$ region. The reconnected junction may capture this $p_T$ increasement and make the CR related $\Lambda_c^+$ shifted to larger characteristic  $p_T$. The difference between the coalescence hadronization model and the color reconnection model on the flavor associated multiplicity dependence of baryon to meson ratios can be further tested with more precise experimental data in the future.

	\section{Summary}
	\label{sec:summary}
	This study systematically investigates the multiplicity dependent hadron production with different flavors in proton proton collisions at $\sqrt{s}=13$ TeV employing the string-melting AMPT model coupled with the PYTHIA8 initial conditions which includes final-state interaction effects and the coalescence hadronization model. We find that the results from the AMPT calculation closely align with those of the color reconnection model and both models describe the inclusive hadron productions reasonably. 
 
    Additionally, we observe that the final state parton stage evolutions, in conjunction with the coalescence process, will lead to a pronounced multiplicity dependence for the baryon to meson ratios, displaying a clear flavor hierarchy. The color reconnection model also predicts a similar multiplicity dependence for the baryon to meson ratio, although it does not clearly delineate the ordering between $p/\pi$ and $\Lambda/K_{S}$. 
    
    The multiplicity induced modifications to the $p_T$ shape of the baryon to meson ratio in AMPT model with different quark components expose the flavor related medium response effects in high energy pp collisions. We think that the discrepancy in the calculations for the flavor hierarchy in the baryon to meson ratio and $p_T$ shape of baryon to meson ratio between AMPT and the color reconnection model can be important to the distinguish the hadronization mechanism at play in high multiplicity pp collisions. 
    
    The study in this work underscores the importance of carrying out multiplicity dependent studies alongside analyzing the flavor hierarchy patterns. Such an approach is essential for gaining insights to understand the collectivity like effects observed in small systems.

	\begin{acknowledgments}
		We would like to thank Xiaoming Zhang, Ziwei Lin for helpful discussions. This work was Supported by the National Natural Science Foundation of China under  Grant No.12205259, No.12147101 and the Fundamental Research Funds for the Central Universities, China University of Geosciences(Wuhan) with No. G1323523064.
		
	\end{acknowledgments}

	\bibliography{reference}

\begin{thebibliography}{65}%
\makeatletter
\providecommand \@ifxundefined [1]{%
 \@ifx{#1\undefined}
}%
\providecommand \@ifnum [1]{%
 \ifnum #1\expandafter \@firstoftwo
 \else \expandafter \@secondoftwo
 \fi
}%
\providecommand \@ifx [1]{%
 \ifx #1\expandafter \@firstoftwo
 \else \expandafter \@secondoftwo
 \fi
}%
\providecommand \natexlab [1]{#1}%
\providecommand \enquote  [1]{``#1''}%
\providecommand \bibnamefont  [1]{#1}%
\providecommand \bibfnamefont [1]{#1}%
\providecommand \citenamefont [1]{#1}%
\providecommand \href@noop [0]{\@secondoftwo}%
\providecommand \href [0]{\begingroup \@sanitize@url \@href}%
\providecommand \@href[1]{\@@startlink{#1}\@@href}%
\providecommand \@@href[1]{\endgroup#1\@@endlink}%
\providecommand \@sanitize@url [0]{\catcode `\\12\catcode `\$12\catcode
  `\&12\catcode `\#12\catcode `\^12\catcode `\_12\catcode `\%12\relax}%
\providecommand \@@startlink[1]{}%
\providecommand \@@endlink[0]{}%
\providecommand \url  [0]{\begingroup\@sanitize@url \@url }%
\providecommand \@url [1]{\endgroup\@href {#1}{\urlprefix }}%
\providecommand \urlprefix  [0]{URL }%
\providecommand \Eprint [0]{\href }%
\providecommand \doibase [0]{http://dx.doi.org/}%
\providecommand \selectlanguage [0]{\@gobble}%
\providecommand \bibinfo  [0]{\@secondoftwo}%
\providecommand \bibfield  [0]{\@secondoftwo}%
\providecommand \translation [1]{[#1]}%
\providecommand \BibitemOpen [0]{}%
\providecommand \bibitemStop [0]{}%
\providecommand \bibitemNoStop [0]{.\EOS\space}%
\providecommand \EOS [0]{\spacefactor3000\relax}%
\providecommand \BibitemShut  [1]{\csname bibitem#1\endcsname}%
\let\auto@bib@innerbib\@empty
\bibitem [{\citenamefont {Broniowski}\ \emph {et~al.}(2008)\citenamefont
  {Broniowski}, \citenamefont {Chojnacki}, \citenamefont {Florkowski},\ and\
  \citenamefont {Kisiel}}]{Broniowski:2008vp}%
  \BibitemOpen
  \bibfield  {author} {\bibinfo {author} {\bibfnamefont {W.}~\bibnamefont
  {Broniowski}}, \bibinfo {author} {\bibfnamefont {M.}~\bibnamefont
  {Chojnacki}}, \bibinfo {author} {\bibfnamefont {W.}~\bibnamefont
  {Florkowski}}, \ and\ \bibinfo {author} {\bibfnamefont {A.}~\bibnamefont
  {Kisiel}},\ }\href {\doibase 10.1103/PhysRevLett.101.022301} {\bibfield
  {journal} {\bibinfo  {journal} {Phys. Rev. Lett.}\ }\textbf {\bibinfo
  {volume} {101}},\ \bibinfo {pages} {022301} (\bibinfo {year} {2008})},\
  \Eprint {http://arxiv.org/abs/0801.4361} {arXiv:0801.4361 [nucl-th]}
  \BibitemShut {NoStop}%
\bibitem [{\citenamefont {Busza}\ \emph {et~al.}(2018)\citenamefont {Busza},
  \citenamefont {Rajagopal},\ and\ \citenamefont {van~der
  Schee}}]{Busza:2018rrf}%
  \BibitemOpen
  \bibfield  {author} {\bibinfo {author} {\bibfnamefont {W.}~\bibnamefont
  {Busza}}, \bibinfo {author} {\bibfnamefont {K.}~\bibnamefont {Rajagopal}}, \
  and\ \bibinfo {author} {\bibfnamefont {W.}~\bibnamefont {van~der Schee}},\
  }\href {\doibase 10.1146/annurev-nucl-101917-020852} {\bibfield  {journal}
  {\bibinfo  {journal} {Ann. Rev. Nucl. Part. Sci.}\ }\textbf {\bibinfo
  {volume} {68}},\ \bibinfo {pages} {339} (\bibinfo {year} {2018})},\ \Eprint
  {http://arxiv.org/abs/1802.04801} {arXiv:1802.04801 [hep-ph]} \BibitemShut
  {NoStop}%
\bibitem [{\citenamefont {Elfner}\ and\ \citenamefont
  {M\"uller}(2023)}]{Elfner:2022iae}%
  \BibitemOpen
  \bibfield  {author} {\bibinfo {author} {\bibfnamefont {H.}~\bibnamefont
  {Elfner}}\ and\ \bibinfo {author} {\bibfnamefont {B.}~\bibnamefont
  {M\"uller}},\ }\href {\doibase 10.1088/1361-6471/ace824} {\bibfield
  {journal} {\bibinfo  {journal} {J. Phys. G}\ }\textbf {\bibinfo {volume}
  {50}},\ \bibinfo {pages} {103001} (\bibinfo {year} {2023})},\ \Eprint
  {http://arxiv.org/abs/2210.12056} {arXiv:2210.12056 [nucl-th]} \BibitemShut
  {NoStop}%
\bibitem [{\citenamefont {Harris}\ and\ \citenamefont
  {M\"uller}(2023)}]{Harris:2023tti}%
  \BibitemOpen
  \bibfield  {author} {\bibinfo {author} {\bibfnamefont {J.~W.}\ \bibnamefont
  {Harris}}\ and\ \bibinfo {author} {\bibfnamefont {B.}~\bibnamefont
  {M\"uller}},\ }\href@noop {} {\  (\bibinfo {year} {2023})},\ \Eprint
  {http://arxiv.org/abs/2308.05743} {arXiv:2308.05743 [hep-ph]} \BibitemShut
  {NoStop}%
\bibitem [{\citenamefont {Yu}\ \emph {et~al.}(2023)\citenamefont {Yu},
  \citenamefont {Dingwei},\ and\ \citenamefont
  {Xiaofeng}}]{ZHANG2023Experimental}%
  \BibitemOpen
  \bibfield  {author} {\bibinfo {author} {\bibfnamefont {Z.}~\bibnamefont
  {Yu}}, \bibinfo {author} {\bibfnamefont {Z.}~\bibnamefont {Dingwei}}, \ and\
  \bibinfo {author} {\bibfnamefont {L.}~\bibnamefont {Xiaofeng}},\ }\href
  {\doibase 10.11889/j.0253-3219.2023.hjs.46.040001} {\bibfield  {journal}
  {\bibinfo  {journal} {NUCLEAR TECHNIQUES}\ }\textbf {\bibinfo {volume} {46}}
  (\bibinfo {year} {2023}),\
  10.11889/j.0253-3219.2023.hjs.46.040001}\BibitemShut {NoStop}%
\bibitem [{\citenamefont {Adolfsson}\ \emph {et~al.}(2020)\citenamefont
  {Adolfsson} \emph {et~al.}}]{Adolfsson:2020dhm}%
  \BibitemOpen
  \bibfield  {author} {\bibinfo {author} {\bibfnamefont {J.}~\bibnamefont
  {Adolfsson}} \emph {et~al.},\ }\href {\doibase
  10.1140/epja/s10050-020-00270-1} {\bibfield  {journal} {\bibinfo  {journal}
  {Eur. Phys. J. A}\ }\textbf {\bibinfo {volume} {56}},\ \bibinfo {pages} {288}
  (\bibinfo {year} {2020})},\ \Eprint {http://arxiv.org/abs/2003.10997}
  {arXiv:2003.10997 [hep-ph]} \BibitemShut {NoStop}%
\bibitem [{\citenamefont {Khachatryan}\ \emph {et~al.}(2010)\citenamefont
  {Khachatryan} \emph {et~al.}}]{CMS:2010ifv}%
  \BibitemOpen
  \bibfield  {author} {\bibinfo {author} {\bibfnamefont {V.}~\bibnamefont
  {Khachatryan}} \emph {et~al.} (\bibinfo {collaboration} {CMS}),\ }\href
  {\doibase 10.1007/JHEP09(2010)091} {\bibfield  {journal} {\bibinfo  {journal}
  {JHEP}\ }\textbf {\bibinfo {volume} {09}},\ \bibinfo {pages} {091} (\bibinfo
  {year} {2010})},\ \Eprint {http://arxiv.org/abs/1009.4122} {arXiv:1009.4122
  [hep-ex]} \BibitemShut {NoStop}%
\bibitem [{\citenamefont {Khachatryan}\ \emph {et~al.}(2017)\citenamefont
  {Khachatryan} \emph {et~al.}}]{CMS:2016fnw}%
  \BibitemOpen
  \bibfield  {author} {\bibinfo {author} {\bibfnamefont {V.}~\bibnamefont
  {Khachatryan}} \emph {et~al.} (\bibinfo {collaboration} {CMS}),\ }\href
  {\doibase 10.1016/j.physletb.2016.12.009} {\bibfield  {journal} {\bibinfo
  {journal} {Phys. Lett. B}\ }\textbf {\bibinfo {volume} {765}},\ \bibinfo
  {pages} {193} (\bibinfo {year} {2017})},\ \Eprint
  {http://arxiv.org/abs/1606.06198} {arXiv:1606.06198 [nucl-ex]} \BibitemShut
  {NoStop}%
\bibitem [{\citenamefont {Adam}\ \emph
  {et~al.}(2017{\natexlab{a}})\citenamefont {Adam} \emph
  {et~al.}}]{ALICE:2016fzo}%
  \BibitemOpen
  \bibfield  {author} {\bibinfo {author} {\bibfnamefont {J.}~\bibnamefont
  {Adam}} \emph {et~al.} (\bibinfo {collaboration} {ALICE}),\ }\href {\doibase
  10.1038/nphys4111} {\bibfield  {journal} {\bibinfo  {journal} {Nature Phys.}\
  }\textbf {\bibinfo {volume} {13}},\ \bibinfo {pages} {535} (\bibinfo {year}
  {2017}{\natexlab{a}})},\ \Eprint {http://arxiv.org/abs/1606.07424}
  {arXiv:1606.07424 [nucl-ex]} \BibitemShut {NoStop}%
\bibitem [{\citenamefont {Adam}\ \emph {et~al.}(2015)\citenamefont {Adam} \emph
  {et~al.}}]{ALICE:2015ial}%
  \BibitemOpen
  \bibfield  {author} {\bibinfo {author} {\bibfnamefont {J.}~\bibnamefont
  {Adam}} \emph {et~al.} (\bibinfo {collaboration} {ALICE}),\ }\href {\doibase
  10.1140/epjc/s10052-015-3422-9} {\bibfield  {journal} {\bibinfo  {journal}
  {Eur. Phys. J. C}\ }\textbf {\bibinfo {volume} {75}},\ \bibinfo {pages} {226}
  (\bibinfo {year} {2015})},\ \Eprint {http://arxiv.org/abs/1504.00024}
  {arXiv:1504.00024 [nucl-ex]} \BibitemShut {NoStop}%
\bibitem [{\citenamefont {Acharya}\ \emph {et~al.}(2019)\citenamefont {Acharya}
  \emph {et~al.}}]{ALICE:2018pal}%
  \BibitemOpen
  \bibfield  {author} {\bibinfo {author} {\bibfnamefont {S.}~\bibnamefont
  {Acharya}} \emph {et~al.} (\bibinfo {collaboration} {ALICE}),\ }\href
  {\doibase 10.1103/PhysRevC.99.024906} {\bibfield  {journal} {\bibinfo
  {journal} {Phys. Rev. C}\ }\textbf {\bibinfo {volume} {99}},\ \bibinfo
  {pages} {024906} (\bibinfo {year} {2019})},\ \Eprint
  {http://arxiv.org/abs/1807.11321} {arXiv:1807.11321 [nucl-ex]} \BibitemShut
  {NoStop}%
\bibitem [{\citenamefont {Shou}\ \emph {et~al.}(2024)\citenamefont {Shou} \emph
  {et~al.}}]{Shou:2024uga}%
  \BibitemOpen
  \bibfield  {author} {\bibinfo {author} {\bibfnamefont {Q.-Y.}\ \bibnamefont
  {Shou}} \emph {et~al.},\ }\href {\doibase 10.1007/s41365-024-01583-2}
  {\bibfield  {journal} {\bibinfo  {journal} {Nucl. Sci. Tech.}\ }\textbf
  {\bibinfo {volume} {35}},\ \bibinfo {pages} {219} (\bibinfo {year} {2024})},\
  \Eprint {http://arxiv.org/abs/2409.17964} {arXiv:2409.17964 [nucl-ex]}
  \BibitemShut {NoStop}%
\bibitem [{\citenamefont {Nagle}\ and\ \citenamefont
  {Zajc}(2018)}]{Nagle:2018nvi}%
  \BibitemOpen
  \bibfield  {author} {\bibinfo {author} {\bibfnamefont {J.~L.}\ \bibnamefont
  {Nagle}}\ and\ \bibinfo {author} {\bibfnamefont {W.~A.}\ \bibnamefont
  {Zajc}},\ }\href {\doibase 10.1146/annurev-nucl-101916-123209} {\bibfield
  {journal} {\bibinfo  {journal} {Ann. Rev. Nucl. Part. Sci.}\ }\textbf
  {\bibinfo {volume} {68}},\ \bibinfo {pages} {211} (\bibinfo {year} {2018})},\
  \Eprint {http://arxiv.org/abs/1801.03477} {arXiv:1801.03477 [nucl-ex]}
  \BibitemShut {NoStop}%
\bibitem [{\citenamefont {Noronha}\ \emph {et~al.}(2024)\citenamefont
  {Noronha}, \citenamefont {Schenke}, \citenamefont {Shen},\ and\ \citenamefont
  {Zhao}}]{Noronha:2024dtq}%
  \BibitemOpen
  \bibfield  {author} {\bibinfo {author} {\bibfnamefont {J.}~\bibnamefont
  {Noronha}}, \bibinfo {author} {\bibfnamefont {B.}~\bibnamefont {Schenke}},
  \bibinfo {author} {\bibfnamefont {C.}~\bibnamefont {Shen}}, \ and\ \bibinfo
  {author} {\bibfnamefont {W.}~\bibnamefont {Zhao}}\ }(\bibinfo {year} {2024})\
  \Eprint {http://arxiv.org/abs/2401.09208} {arXiv:2401.09208 [nucl-th]}
  \BibitemShut {NoStop}%
\bibitem [{\citenamefont {Adam}\ \emph
  {et~al.}(2017{\natexlab{b}})\citenamefont {Adam} \emph
  {et~al.}}]{ALICE:2017jyt}%
  \BibitemOpen
  \bibfield  {author} {\bibinfo {author} {\bibfnamefont {J.}~\bibnamefont
  {Adam}} \emph {et~al.} (\bibinfo {collaboration} {ALICE}),\ }\href {\doibase
  10.1038/nphys4111} {\bibfield  {journal} {\bibinfo  {journal} {Nature Phys.}\
  }\textbf {\bibinfo {volume} {13}},\ \bibinfo {pages} {535} (\bibinfo {year}
  {2017}{\natexlab{b}})},\ \Eprint {http://arxiv.org/abs/1606.07424}
  {arXiv:1606.07424 [nucl-ex]} \BibitemShut {NoStop}%
\bibitem [{\citenamefont {Acharya}\ \emph
  {et~al.}(2020{\natexlab{a}})\citenamefont {Acharya} \emph
  {et~al.}}]{ALICE:2020nkc}%
  \BibitemOpen
  \bibfield  {author} {\bibinfo {author} {\bibfnamefont {S.}~\bibnamefont
  {Acharya}} \emph {et~al.} (\bibinfo {collaboration} {ALICE}),\ }\href
  {\doibase 10.1140/epjc/s10052-020-8125-1} {\bibfield  {journal} {\bibinfo
  {journal} {Eur. Phys. J. C}\ }\textbf {\bibinfo {volume} {80}},\ \bibinfo
  {pages} {693} (\bibinfo {year} {2020}{\natexlab{a}})},\ \Eprint
  {http://arxiv.org/abs/2003.02394} {arXiv:2003.02394 [nucl-ex]} \BibitemShut
  {NoStop}%
\bibitem [{\citenamefont {Acharya}\ \emph {et~al.}(2024)\citenamefont {Acharya}
  \emph {et~al.}}]{ALICE:2023edr}%
  \BibitemOpen
  \bibfield  {author} {\bibinfo {author} {\bibfnamefont {S.}~\bibnamefont
  {Acharya}} \emph {et~al.} (\bibinfo {collaboration} {ALICE}),\ }\href
  {\doibase 10.1103/PhysRevC.109.014911} {\bibfield  {journal} {\bibinfo
  {journal} {Phys. Rev. C}\ }\textbf {\bibinfo {volume} {109}},\ \bibinfo
  {pages} {014911} (\bibinfo {year} {2024})},\ \Eprint
  {http://arxiv.org/abs/2308.16115} {arXiv:2308.16115 [nucl-ex]} \BibitemShut
  {NoStop}%
\bibitem [{\citenamefont {Acharya}\ \emph
  {et~al.}(2023{\natexlab{a}})\citenamefont {Acharya} \emph
  {et~al.}}]{ALICE:2022imr}%
  \BibitemOpen
  \bibfield  {author} {\bibinfo {author} {\bibfnamefont {S.}~\bibnamefont
  {Acharya}} \emph {et~al.} (\bibinfo {collaboration} {ALICE}),\ }\href
  {\doibase 10.1016/j.physletb.2023.137730} {\bibfield  {journal} {\bibinfo
  {journal} {Phys. Lett. B}\ }\textbf {\bibinfo {volume} {845}},\ \bibinfo
  {pages} {137730} (\bibinfo {year} {2023}{\natexlab{a}})},\ \Eprint
  {http://arxiv.org/abs/2204.10210} {arXiv:2204.10210 [nucl-ex]} \BibitemShut
  {NoStop}%
\bibitem [{\citenamefont {Acharya}\ \emph
  {et~al.}(2020{\natexlab{b}})\citenamefont {Acharya} \emph
  {et~al.}}]{ALICE:2019avo}%
  \BibitemOpen
  \bibfield  {author} {\bibinfo {author} {\bibfnamefont {S.}~\bibnamefont
  {Acharya}} \emph {et~al.} (\bibinfo {collaboration} {ALICE}),\ }\href
  {\doibase 10.1140/epjc/s10052-020-7673-8} {\bibfield  {journal} {\bibinfo
  {journal} {Eur. Phys. J. C}\ }\textbf {\bibinfo {volume} {80}},\ \bibinfo
  {pages} {167} (\bibinfo {year} {2020}{\natexlab{b}})},\ \Eprint
  {http://arxiv.org/abs/1908.01861} {arXiv:1908.01861 [nucl-ex]} \BibitemShut
  {NoStop}%
\bibitem [{\citenamefont {Acharya}\ \emph
  {et~al.}(2020{\natexlab{c}})\citenamefont {Acharya} \emph
  {et~al.}}]{ALICE:2019etb}%
  \BibitemOpen
  \bibfield  {author} {\bibinfo {author} {\bibfnamefont {S.}~\bibnamefont
  {Acharya}} \emph {et~al.} (\bibinfo {collaboration} {ALICE}),\ }\href
  {\doibase 10.1016/j.physletb.2020.135501} {\bibfield  {journal} {\bibinfo
  {journal} {Phys. Lett. B}\ }\textbf {\bibinfo {volume} {807}},\ \bibinfo
  {pages} {135501} (\bibinfo {year} {2020}{\natexlab{c}})},\ \Eprint
  {http://arxiv.org/abs/1910.14397} {arXiv:1910.14397 [nucl-ex]} \BibitemShut
  {NoStop}%
\bibitem [{\citenamefont {Acharya}\ \emph {et~al.}(2022)\citenamefont {Acharya}
  \emph {et~al.}}]{ALICE:2021npz}%
  \BibitemOpen
  \bibfield  {author} {\bibinfo {author} {\bibfnamefont {S.}~\bibnamefont
  {Acharya}} \emph {et~al.} (\bibinfo {collaboration} {ALICE}),\ }\href
  {\doibase 10.1016/j.physletb.2022.137065} {\bibfield  {journal} {\bibinfo
  {journal} {Phys. Lett. B}\ }\textbf {\bibinfo {volume} {829}},\ \bibinfo
  {pages} {137065} (\bibinfo {year} {2022})},\ \Eprint
  {http://arxiv.org/abs/2111.11948} {arXiv:2111.11948 [nucl-ex]} \BibitemShut
  {NoStop}%
\bibitem [{\citenamefont {Kanakubo}\ \emph {et~al.}(2018)\citenamefont
  {Kanakubo}, \citenamefont {Okai}, \citenamefont {Tachibana},\ and\
  \citenamefont {Hirano}}]{Kanakubo:2018vkl}%
  \BibitemOpen
  \bibfield  {author} {\bibinfo {author} {\bibfnamefont {Y.}~\bibnamefont
  {Kanakubo}}, \bibinfo {author} {\bibfnamefont {M.}~\bibnamefont {Okai}},
  \bibinfo {author} {\bibfnamefont {Y.}~\bibnamefont {Tachibana}}, \ and\
  \bibinfo {author} {\bibfnamefont {T.}~\bibnamefont {Hirano}},\ }\href
  {\doibase 10.1093/ptep/pty129} {\bibfield  {journal} {\bibinfo  {journal}
  {PTEP}\ }\textbf {\bibinfo {volume} {2018}},\ \bibinfo {pages} {121D01}
  (\bibinfo {year} {2018})},\ \Eprint {http://arxiv.org/abs/1806.10329}
  {arXiv:1806.10329 [nucl-th]} \BibitemShut {NoStop}%
\bibitem [{\citenamefont {Kanakubo}\ \emph {et~al.}(2020)\citenamefont
  {Kanakubo}, \citenamefont {Tachibana},\ and\ \citenamefont
  {Hirano}}]{Kanakubo:2019ogh}%
  \BibitemOpen
  \bibfield  {author} {\bibinfo {author} {\bibfnamefont {Y.}~\bibnamefont
  {Kanakubo}}, \bibinfo {author} {\bibfnamefont {Y.}~\bibnamefont {Tachibana}},
  \ and\ \bibinfo {author} {\bibfnamefont {T.}~\bibnamefont {Hirano}},\ }\href
  {\doibase 10.1103/PhysRevC.101.024912} {\bibfield  {journal} {\bibinfo
  {journal} {Phys. Rev. C}\ }\textbf {\bibinfo {volume} {101}},\ \bibinfo
  {pages} {024912} (\bibinfo {year} {2020})},\ \Eprint
  {http://arxiv.org/abs/1910.10556} {arXiv:1910.10556 [nucl-th]} \BibitemShut
  {NoStop}%
\bibitem [{\citenamefont {Vislavicius}\ and\ \citenamefont
  {Kalweit}(2016)}]{Vislavicius:2016rwi}%
  \BibitemOpen
  \bibfield  {author} {\bibinfo {author} {\bibfnamefont {V.}~\bibnamefont
  {Vislavicius}}\ and\ \bibinfo {author} {\bibfnamefont {A.}~\bibnamefont
  {Kalweit}},\ }\href@noop {} {\  (\bibinfo {year} {2016})},\ \Eprint
  {http://arxiv.org/abs/1610.03001} {arXiv:1610.03001 [nucl-ex]} \BibitemShut
  {NoStop}%
\bibitem [{\citenamefont {Zhao}\ \emph {et~al.}(2018)\citenamefont {Zhao},
  \citenamefont {Zhou}, \citenamefont {Xu}, \citenamefont {Deng},\ and\
  \citenamefont {Song}}]{Zhao:2017rgg}%
  \BibitemOpen
  \bibfield  {author} {\bibinfo {author} {\bibfnamefont {W.}~\bibnamefont
  {Zhao}}, \bibinfo {author} {\bibfnamefont {Y.}~\bibnamefont {Zhou}}, \bibinfo
  {author} {\bibfnamefont {H.}~\bibnamefont {Xu}}, \bibinfo {author}
  {\bibfnamefont {W.}~\bibnamefont {Deng}}, \ and\ \bibinfo {author}
  {\bibfnamefont {H.}~\bibnamefont {Song}},\ }\href {\doibase
  10.1016/j.physletb.2018.03.022} {\bibfield  {journal} {\bibinfo  {journal}
  {Phys. Lett. B}\ }\textbf {\bibinfo {volume} {780}},\ \bibinfo {pages} {495}
  (\bibinfo {year} {2018})},\ \Eprint {http://arxiv.org/abs/1801.00271}
  {arXiv:1801.00271 [nucl-th]} \BibitemShut {NoStop}%
\bibitem [{\citenamefont {Zhao}\ \emph {et~al.}(2020)\citenamefont {Zhao},
  \citenamefont {Zhou}, \citenamefont {Murase},\ and\ \citenamefont
  {Song}}]{Zhao:2020pty}%
  \BibitemOpen
  \bibfield  {author} {\bibinfo {author} {\bibfnamefont {W.}~\bibnamefont
  {Zhao}}, \bibinfo {author} {\bibfnamefont {Y.}~\bibnamefont {Zhou}}, \bibinfo
  {author} {\bibfnamefont {K.}~\bibnamefont {Murase}}, \ and\ \bibinfo {author}
  {\bibfnamefont {H.}~\bibnamefont {Song}},\ }\href {\doibase
  10.1140/epjc/s10052-020-8376-x} {\bibfield  {journal} {\bibinfo  {journal}
  {Eur. Phys. J. C}\ }\textbf {\bibinfo {volume} {80}},\ \bibinfo {pages} {846}
  (\bibinfo {year} {2020})},\ \Eprint {http://arxiv.org/abs/2001.06742}
  {arXiv:2001.06742 [nucl-th]} \BibitemShut {NoStop}%
\bibitem [{\citenamefont {Dong}\ \emph {et~al.}(2024)\citenamefont {Dong},
  \citenamefont {Yu}, \citenamefont {Ping}, \citenamefont {Wu}, \citenamefont
  {Wang}, \citenamefont {Huang},\ and\ \citenamefont {Lin}}]{Dong:2023zbu}%
  \BibitemOpen
  \bibfield  {author} {\bibinfo {author} {\bibfnamefont {W.-J.}\ \bibnamefont
  {Dong}}, \bibinfo {author} {\bibfnamefont {X.-Z.}\ \bibnamefont {Yu}},
  \bibinfo {author} {\bibfnamefont {S.-Y.}\ \bibnamefont {Ping}}, \bibinfo
  {author} {\bibfnamefont {X.-T.}\ \bibnamefont {Wu}}, \bibinfo {author}
  {\bibfnamefont {G.}~\bibnamefont {Wang}}, \bibinfo {author} {\bibfnamefont
  {H.~Z.}\ \bibnamefont {Huang}}, \ and\ \bibinfo {author} {\bibfnamefont
  {Z.-W.}\ \bibnamefont {Lin}},\ }\href {\doibase 10.1007/s41365-024-01464-8}
  {\bibfield  {journal} {\bibinfo  {journal} {Nucl. Sci. Tech.}\ }\textbf
  {\bibinfo {volume} {35}},\ \bibinfo {pages} {120} (\bibinfo {year} {2024})},\
  \Eprint {http://arxiv.org/abs/2306.15160} {arXiv:2306.15160 [hep-ph]}
  \BibitemShut {NoStop}%
\bibitem [{\citenamefont {Tang}\ \emph {et~al.}(2024)\citenamefont {Tang},
  \citenamefont {Zheng}, \citenamefont {Zhang},\ and\ \citenamefont
  {Wan}}]{Tang:2023wcd}%
  \BibitemOpen
  \bibfield  {author} {\bibinfo {author} {\bibfnamefont {S.-Y.}\ \bibnamefont
  {Tang}}, \bibinfo {author} {\bibfnamefont {L.}~\bibnamefont {Zheng}},
  \bibinfo {author} {\bibfnamefont {X.-M.}\ \bibnamefont {Zhang}}, \ and\
  \bibinfo {author} {\bibfnamefont {R.-Z.}\ \bibnamefont {Wan}},\ }\href
  {\doibase 10.1007/s41365-024-01387-4} {\bibfield  {journal} {\bibinfo
  {journal} {Nucl. Sci. Tech.}\ }\textbf {\bibinfo {volume} {35}},\ \bibinfo
  {pages} {32} (\bibinfo {year} {2024})},\ \Eprint
  {http://arxiv.org/abs/2303.06577} {arXiv:2303.06577 [hep-ph]} \BibitemShut
  {NoStop}%
\bibitem [{\citenamefont {Mazeliauskas}\ and\ \citenamefont
  {Vislavicius}(2020)}]{Mazeliauskas:2019ifr}%
  \BibitemOpen
  \bibfield  {author} {\bibinfo {author} {\bibfnamefont {A.}~\bibnamefont
  {Mazeliauskas}}\ and\ \bibinfo {author} {\bibfnamefont {V.}~\bibnamefont
  {Vislavicius}},\ }\href {\doibase 10.1103/PhysRevC.101.014910} {\bibfield
  {journal} {\bibinfo  {journal} {Phys. Rev. C}\ }\textbf {\bibinfo {volume}
  {101}},\ \bibinfo {pages} {014910} (\bibinfo {year} {2020})},\ \Eprint
  {http://arxiv.org/abs/1907.11059} {arXiv:1907.11059 [hep-ph]} \BibitemShut
  {NoStop}%
\bibitem [{\citenamefont {Motornenko}\ \emph {et~al.}(2020)\citenamefont
  {Motornenko}, \citenamefont {Vovchenko}, \citenamefont {Greiner},\ and\
  \citenamefont {Stoecker}}]{Motornenko:2019jha}%
  \BibitemOpen
  \bibfield  {author} {\bibinfo {author} {\bibfnamefont {A.}~\bibnamefont
  {Motornenko}}, \bibinfo {author} {\bibfnamefont {V.}~\bibnamefont
  {Vovchenko}}, \bibinfo {author} {\bibfnamefont {C.}~\bibnamefont {Greiner}},
  \ and\ \bibinfo {author} {\bibfnamefont {H.}~\bibnamefont {Stoecker}},\
  }\href {\doibase 10.1103/PhysRevC.102.024909} {\bibfield  {journal} {\bibinfo
   {journal} {Phys. Rev. C}\ }\textbf {\bibinfo {volume} {102}},\ \bibinfo
  {pages} {024909} (\bibinfo {year} {2020})},\ \Eprint
  {http://arxiv.org/abs/1908.11730} {arXiv:1908.11730 [hep-ph]} \BibitemShut
  {NoStop}%
\bibitem [{\citenamefont {Flor}\ \emph {et~al.}(2022)\citenamefont {Flor},
  \citenamefont {Olinger},\ and\ \citenamefont {Bellwied}}]{Flor:2021olm}%
  \BibitemOpen
  \bibfield  {author} {\bibinfo {author} {\bibfnamefont {F.~A.}\ \bibnamefont
  {Flor}}, \bibinfo {author} {\bibfnamefont {G.}~\bibnamefont {Olinger}}, \
  and\ \bibinfo {author} {\bibfnamefont {R.}~\bibnamefont {Bellwied}},\ }\href
  {\doibase 10.1016/j.physletb.2022.137473} {\bibfield  {journal} {\bibinfo
  {journal} {Phys. Lett. B}\ }\textbf {\bibinfo {volume} {834}},\ \bibinfo
  {pages} {137473} (\bibinfo {year} {2022})},\ \Eprint
  {http://arxiv.org/abs/2109.09843} {arXiv:2109.09843 [nucl-ex]} \BibitemShut
  {NoStop}%
\bibitem [{\citenamefont {B\'\i{}r\'o}\ \emph {et~al.}(2020)\citenamefont
  {B\'\i{}r\'o}, \citenamefont {Barnaf\"oldi},\ and\ \citenamefont
  {Bir\'o}}]{Biro:2020kve}%
  \BibitemOpen
  \bibfield  {author} {\bibinfo {author} {\bibfnamefont {G.}~\bibnamefont
  {B\'\i{}r\'o}}, \bibinfo {author} {\bibfnamefont {G.~G.}\ \bibnamefont
  {Barnaf\"oldi}}, \ and\ \bibinfo {author} {\bibfnamefont {T.~S.}\
  \bibnamefont {Bir\'o}},\ }\href {\doibase 10.1088/1361-6471/ab8dcb}
  {\bibfield  {journal} {\bibinfo  {journal} {J. Phys. G}\ }\textbf {\bibinfo
  {volume} {47}},\ \bibinfo {pages} {105002} (\bibinfo {year} {2020})},\
  \Eprint {http://arxiv.org/abs/2003.03278} {arXiv:2003.03278 [hep-ph]}
  \BibitemShut {NoStop}%
\bibitem [{\citenamefont {Bierlich}\ and\ \citenamefont
  {Christiansen}(2015)}]{Bierlich:2015rha}%
  \BibitemOpen
  \bibfield  {author} {\bibinfo {author} {\bibfnamefont {C.}~\bibnamefont
  {Bierlich}}\ and\ \bibinfo {author} {\bibfnamefont {J.~R.}\ \bibnamefont
  {Christiansen}},\ }\href {\doibase 10.1103/PhysRevD.92.094010} {\bibfield
  {journal} {\bibinfo  {journal} {Phys. Rev. D}\ }\textbf {\bibinfo {volume}
  {92}},\ \bibinfo {pages} {094010} (\bibinfo {year} {2015})},\ \Eprint
  {http://arxiv.org/abs/1507.02091} {arXiv:1507.02091 [hep-ph]} \BibitemShut
  {NoStop}%
\bibitem [{\citenamefont {Bierlich}\ \emph {et~al.}(2016)\citenamefont
  {Bierlich}, \citenamefont {Gustafson},\ and\ \citenamefont
  {L\"onnblad}}]{Bierlich:2016vgw}%
  \BibitemOpen
  \bibfield  {author} {\bibinfo {author} {\bibfnamefont {C.}~\bibnamefont
  {Bierlich}}, \bibinfo {author} {\bibfnamefont {G.}~\bibnamefont {Gustafson}},
  \ and\ \bibinfo {author} {\bibfnamefont {L.}~\bibnamefont {L\"onnblad}},\
  }\href@noop {} {\  (\bibinfo {year} {2016})},\ \Eprint
  {http://arxiv.org/abs/1612.05132} {arXiv:1612.05132 [hep-ph]} \BibitemShut
  {NoStop}%
\bibitem [{\citenamefont {Bierlich}\ \emph {et~al.}(2018)\citenamefont
  {Bierlich}, \citenamefont {Gustafson},\ and\ \citenamefont
  {L\"onnblad}}]{Bierlich:2017vhg}%
  \BibitemOpen
  \bibfield  {author} {\bibinfo {author} {\bibfnamefont {C.}~\bibnamefont
  {Bierlich}}, \bibinfo {author} {\bibfnamefont {G.}~\bibnamefont {Gustafson}},
  \ and\ \bibinfo {author} {\bibfnamefont {L.}~\bibnamefont {L\"onnblad}},\
  }\href {\doibase 10.1016/j.physletb.2018.01.069} {\bibfield  {journal}
  {\bibinfo  {journal} {Phys. Lett. B}\ }\textbf {\bibinfo {volume} {779}},\
  \bibinfo {pages} {58} (\bibinfo {year} {2018})},\ \Eprint
  {http://arxiv.org/abs/1710.09725} {arXiv:1710.09725 [hep-ph]} \BibitemShut
  {NoStop}%
\bibitem [{\citenamefont {Bierlich}\ \emph
  {et~al.}(2021{\natexlab{a}})\citenamefont {Bierlich}, \citenamefont
  {Chakraborty}, \citenamefont {Gustafson},\ and\ \citenamefont
  {L\"onnblad}}]{Bierlich:2020naj}%
  \BibitemOpen
  \bibfield  {author} {\bibinfo {author} {\bibfnamefont {C.}~\bibnamefont
  {Bierlich}}, \bibinfo {author} {\bibfnamefont {S.}~\bibnamefont
  {Chakraborty}}, \bibinfo {author} {\bibfnamefont {G.}~\bibnamefont
  {Gustafson}}, \ and\ \bibinfo {author} {\bibfnamefont {L.}~\bibnamefont
  {L\"onnblad}},\ }\href {\doibase 10.1007/JHEP03(2021)270} {\bibfield
  {journal} {\bibinfo  {journal} {JHEP}\ }\textbf {\bibinfo {volume} {03}},\
  \bibinfo {pages} {270} (\bibinfo {year} {2021}{\natexlab{a}})},\ \Eprint
  {http://arxiv.org/abs/2010.07595} {arXiv:2010.07595 [hep-ph]} \BibitemShut
  {NoStop}%
\bibitem [{\citenamefont {Bierlich}\ \emph
  {et~al.}(2021{\natexlab{b}})\citenamefont {Bierlich}, \citenamefont
  {Sj\"ostrand},\ and\ \citenamefont {Utheim}}]{Bierlich:2021poz}%
  \BibitemOpen
  \bibfield  {author} {\bibinfo {author} {\bibfnamefont {C.}~\bibnamefont
  {Bierlich}}, \bibinfo {author} {\bibfnamefont {T.}~\bibnamefont
  {Sj\"ostrand}}, \ and\ \bibinfo {author} {\bibfnamefont {M.}~\bibnamefont
  {Utheim}},\ }\href {\doibase 10.1140/epja/s10050-021-00543-3} {\bibfield
  {journal} {\bibinfo  {journal} {Eur. Phys. J. A}\ }\textbf {\bibinfo {volume}
  {57}},\ \bibinfo {pages} {227} (\bibinfo {year} {2021}{\natexlab{b}})},\
  \Eprint {http://arxiv.org/abs/2103.09665} {arXiv:2103.09665 [hep-ph]}
  \BibitemShut {NoStop}%
\bibitem [{\citenamefont {Bierlich}(2024)}]{Bierlich:2024odg}%
  \BibitemOpen
  \bibfield  {author} {\bibinfo {author} {\bibfnamefont {C.}~\bibnamefont
  {Bierlich}},\ }\href {\doibase 10.3390/universe10010046} {\bibfield
  {journal} {\bibinfo  {journal} {Universe}\ }\textbf {\bibinfo {volume}
  {10}},\ \bibinfo {pages} {46} (\bibinfo {year} {2024})},\ \Eprint
  {http://arxiv.org/abs/2401.07585} {arXiv:2401.07585 [hep-ph]} \BibitemShut
  {NoStop}%
\bibitem [{\citenamefont {Ortiz~Velasquez}\ \emph {et~al.}(2013)\citenamefont
  {Ortiz~Velasquez}, \citenamefont {Christiansen}, \citenamefont
  {Cuautle~Flores}, \citenamefont {Maldonado~Cervantes},\ and\ \citenamefont
  {Pai\'c}}]{OrtizVelasquez:2013ofg}%
  \BibitemOpen
  \bibfield  {author} {\bibinfo {author} {\bibfnamefont {A.}~\bibnamefont
  {Ortiz~Velasquez}}, \bibinfo {author} {\bibfnamefont {P.}~\bibnamefont
  {Christiansen}}, \bibinfo {author} {\bibfnamefont {E.}~\bibnamefont
  {Cuautle~Flores}}, \bibinfo {author} {\bibfnamefont {I.}~\bibnamefont
  {Maldonado~Cervantes}}, \ and\ \bibinfo {author} {\bibfnamefont
  {G.}~\bibnamefont {Pai\'c}},\ }\href {\doibase
  10.1103/PhysRevLett.111.042001} {\bibfield  {journal} {\bibinfo  {journal}
  {Phys. Rev. Lett.}\ }\textbf {\bibinfo {volume} {111}},\ \bibinfo {pages}
  {042001} (\bibinfo {year} {2013})},\ \Eprint {http://arxiv.org/abs/1303.6326}
  {arXiv:1303.6326 [hep-ph]} \BibitemShut {NoStop}%
\bibitem [{\citenamefont {Bierlich}\ \emph {et~al.}(2015)\citenamefont
  {Bierlich}, \citenamefont {Gustafson}, \citenamefont {L\"onnblad},\ and\
  \citenamefont {Tarasov}}]{Bierlich:2014xba}%
  \BibitemOpen
  \bibfield  {author} {\bibinfo {author} {\bibfnamefont {C.}~\bibnamefont
  {Bierlich}}, \bibinfo {author} {\bibfnamefont {G.}~\bibnamefont {Gustafson}},
  \bibinfo {author} {\bibfnamefont {L.}~\bibnamefont {L\"onnblad}}, \ and\
  \bibinfo {author} {\bibfnamefont {A.}~\bibnamefont {Tarasov}},\ }\href
  {\doibase 10.1007/JHEP03(2015)148} {\bibfield  {journal} {\bibinfo  {journal}
  {JHEP}\ }\textbf {\bibinfo {volume} {03}},\ \bibinfo {pages} {148} (\bibinfo
  {year} {2015})},\ \Eprint {http://arxiv.org/abs/1412.6259} {arXiv:1412.6259
  [hep-ph]} \BibitemShut {NoStop}%
\bibitem [{\citenamefont {Acharya}\ \emph
  {et~al.}(2023{\natexlab{b}})\citenamefont {Acharya} \emph
  {et~al.}}]{ALICE:2023sgl}%
  \BibitemOpen
  \bibfield  {author} {\bibinfo {author} {\bibfnamefont {S.}~\bibnamefont
  {Acharya}} \emph {et~al.} (\bibinfo {collaboration} {ALICE}),\ }\href
  {\doibase 10.1007/JHEP12(2023)086} {\bibfield  {journal} {\bibinfo  {journal}
  {JHEP}\ }\textbf {\bibinfo {volume} {12}},\ \bibinfo {pages} {086} (\bibinfo
  {year} {2023}{\natexlab{b}})},\ \Eprint {http://arxiv.org/abs/2308.04877}
  {arXiv:2308.04877 [hep-ex]} \BibitemShut {NoStop}%
\bibitem [{\citenamefont {Acharya}\ \emph
  {et~al.}(2023{\natexlab{c}})\citenamefont {Acharya} \emph
  {et~al.}}]{ALICE:2023wbx}%
  \BibitemOpen
  \bibfield  {author} {\bibinfo {author} {\bibfnamefont {S.}~\bibnamefont
  {Acharya}} \emph {et~al.} (\bibinfo {collaboration} {ALICE}),\ }\href
  {\doibase 10.1103/PhysRevD.108.112003} {\bibfield  {journal} {\bibinfo
  {journal} {Phys. Rev. D}\ }\textbf {\bibinfo {volume} {108}},\ \bibinfo
  {pages} {112003} (\bibinfo {year} {2023}{\natexlab{c}})},\ \Eprint
  {http://arxiv.org/abs/2308.04873} {arXiv:2308.04873 [hep-ex]} \BibitemShut
  {NoStop}%
\bibitem [{\citenamefont {Kniehl}\ \emph {et~al.}(2020)\citenamefont {Kniehl},
  \citenamefont {Kramer}, \citenamefont {Schienbein},\ and\ \citenamefont
  {Spiesberger}}]{Kniehl:2020szu}%
  \BibitemOpen
  \bibfield  {author} {\bibinfo {author} {\bibfnamefont {B.~A.}\ \bibnamefont
  {Kniehl}}, \bibinfo {author} {\bibfnamefont {G.}~\bibnamefont {Kramer}},
  \bibinfo {author} {\bibfnamefont {I.}~\bibnamefont {Schienbein}}, \ and\
  \bibinfo {author} {\bibfnamefont {H.}~\bibnamefont {Spiesberger}},\ }\href
  {\doibase 10.1103/PhysRevD.101.114021} {\bibfield  {journal} {\bibinfo
  {journal} {Phys. Rev. D}\ }\textbf {\bibinfo {volume} {101}},\ \bibinfo
  {pages} {114021} (\bibinfo {year} {2020})},\ \Eprint
  {http://arxiv.org/abs/2004.04213} {arXiv:2004.04213 [hep-ph]} \BibitemShut
  {NoStop}%
\bibitem [{\citenamefont {Tumasyan}\ \emph {et~al.}(2024)\citenamefont
  {Tumasyan} \emph {et~al.}}]{CMS:2023frs}%
  \BibitemOpen
  \bibfield  {author} {\bibinfo {author} {\bibfnamefont {A.}~\bibnamefont
  {Tumasyan}} \emph {et~al.} (\bibinfo {collaboration} {CMS}),\ }\href
  {\doibase 10.1007/JHEP01(2024)128} {\bibfield  {journal} {\bibinfo  {journal}
  {JHEP}\ }\textbf {\bibinfo {volume} {01}},\ \bibinfo {pages} {128} (\bibinfo
  {year} {2024})},\ \Eprint {http://arxiv.org/abs/2307.11186} {arXiv:2307.11186
  [nucl-ex]} \BibitemShut {NoStop}%
\bibitem [{\citenamefont {Bierlich}\ \emph {et~al.}(2024)\citenamefont
  {Bierlich}, \citenamefont {Gustafson}, \citenamefont {L\"onnblad},\ and\
  \citenamefont {Shah}}]{Bierlich:2023okq}%
  \BibitemOpen
  \bibfield  {author} {\bibinfo {author} {\bibfnamefont {C.}~\bibnamefont
  {Bierlich}}, \bibinfo {author} {\bibfnamefont {G.}~\bibnamefont {Gustafson}},
  \bibinfo {author} {\bibfnamefont {L.}~\bibnamefont {L\"onnblad}}, \ and\
  \bibinfo {author} {\bibfnamefont {H.}~\bibnamefont {Shah}},\ }\href {\doibase
  10.1140/epjc/s10052-024-12593-0} {\bibfield  {journal} {\bibinfo  {journal}
  {Eur. Phys. J. C}\ }\textbf {\bibinfo {volume} {84}},\ \bibinfo {pages} {231}
  (\bibinfo {year} {2024})},\ \Eprint {http://arxiv.org/abs/2309.12452}
  {arXiv:2309.12452 [hep-ph]} \BibitemShut {NoStop}%
\bibitem [{\citenamefont {Song}\ \emph {et~al.}(2018)\citenamefont {Song},
  \citenamefont {Li},\ and\ \citenamefont {Shao}}]{Song:2018tpv}%
  \BibitemOpen
  \bibfield  {author} {\bibinfo {author} {\bibfnamefont {J.}~\bibnamefont
  {Song}}, \bibinfo {author} {\bibfnamefont {H.-h.}\ \bibnamefont {Li}}, \ and\
  \bibinfo {author} {\bibfnamefont {F.-l.}\ \bibnamefont {Shao}},\ }\href
  {\doibase 10.1140/epjc/s10052-018-5817-x} {\bibfield  {journal} {\bibinfo
  {journal} {Eur. Phys. J. C}\ }\textbf {\bibinfo {volume} {78}},\ \bibinfo
  {pages} {344} (\bibinfo {year} {2018})},\ \Eprint
  {http://arxiv.org/abs/1801.09402} {arXiv:1801.09402 [hep-ph]} \BibitemShut
  {NoStop}%
\bibitem [{\citenamefont {Chen}\ and\ \citenamefont {He}(2021)}]{Chen:2020drg}%
  \BibitemOpen
  \bibfield  {author} {\bibinfo {author} {\bibfnamefont {Y.}~\bibnamefont
  {Chen}}\ and\ \bibinfo {author} {\bibfnamefont {M.}~\bibnamefont {He}},\
  }\href {\doibase 10.1016/j.physletb.2021.136144} {\bibfield  {journal}
  {\bibinfo  {journal} {Phys. Lett. B}\ }\textbf {\bibinfo {volume} {815}},\
  \bibinfo {pages} {136144} (\bibinfo {year} {2021})},\ \Eprint
  {http://arxiv.org/abs/2011.14328} {arXiv:2011.14328 [hep-ph]} \BibitemShut
  {NoStop}%
\bibitem [{\citenamefont {He}\ and\ \citenamefont {Rapp}(2019)}]{He:2019tik}%
  \BibitemOpen
  \bibfield  {author} {\bibinfo {author} {\bibfnamefont {M.}~\bibnamefont
  {He}}\ and\ \bibinfo {author} {\bibfnamefont {R.}~\bibnamefont {Rapp}},\
  }\href {\doibase 10.1016/j.physletb.2019.06.004} {\bibfield  {journal}
  {\bibinfo  {journal} {Phys. Lett. B}\ }\textbf {\bibinfo {volume} {795}},\
  \bibinfo {pages} {117} (\bibinfo {year} {2019})},\ \Eprint
  {http://arxiv.org/abs/1902.08889} {arXiv:1902.08889 [nucl-th]} \BibitemShut
  {NoStop}%
\bibitem [{\citenamefont {Minissale}\ \emph {et~al.}(2021)\citenamefont
  {Minissale}, \citenamefont {Plumari},\ and\ \citenamefont
  {Greco}}]{Minissale:2020bif}%
  \BibitemOpen
  \bibfield  {author} {\bibinfo {author} {\bibfnamefont {V.}~\bibnamefont
  {Minissale}}, \bibinfo {author} {\bibfnamefont {S.}~\bibnamefont {Plumari}},
  \ and\ \bibinfo {author} {\bibfnamefont {V.}~\bibnamefont {Greco}},\ }\href
  {\doibase 10.1016/j.physletb.2021.136622} {\bibfield  {journal} {\bibinfo
  {journal} {Phys. Lett. B}\ }\textbf {\bibinfo {volume} {821}},\ \bibinfo
  {pages} {136622} (\bibinfo {year} {2021})},\ \Eprint
  {http://arxiv.org/abs/2012.12001} {arXiv:2012.12001 [hep-ph]} \BibitemShut
  {NoStop}%
\bibitem [{\citenamefont {Zhao}\ \emph {et~al.}(2024)\citenamefont {Zhao},
  \citenamefont {Aichelin}, \citenamefont {Gossiaux},\ and\ \citenamefont
  {Werner}}]{Zhao:2023ucp}%
  \BibitemOpen
  \bibfield  {author} {\bibinfo {author} {\bibfnamefont {J.}~\bibnamefont
  {Zhao}}, \bibinfo {author} {\bibfnamefont {J.}~\bibnamefont {Aichelin}},
  \bibinfo {author} {\bibfnamefont {P.~B.}\ \bibnamefont {Gossiaux}}, \ and\
  \bibinfo {author} {\bibfnamefont {K.}~\bibnamefont {Werner}},\ }\href
  {\doibase 10.1103/PhysRevD.109.054011} {\bibfield  {journal} {\bibinfo
  {journal} {Phys. Rev. D}\ }\textbf {\bibinfo {volume} {109}},\ \bibinfo
  {pages} {054011} (\bibinfo {year} {2024})},\ \Eprint
  {http://arxiv.org/abs/2310.08684} {arXiv:2310.08684 [hep-ph]} \BibitemShut
  {NoStop}%
\bibitem [{\citenamefont {Lin}\ and\ \citenamefont
  {Zheng}(2021)}]{Lin:2021mdn}%
  \BibitemOpen
  \bibfield  {author} {\bibinfo {author} {\bibfnamefont {Z.-W.}\ \bibnamefont
  {Lin}}\ and\ \bibinfo {author} {\bibfnamefont {L.}~\bibnamefont {Zheng}},\
  }\href {\doibase 10.1007/s41365-021-00944-5} {\bibfield  {journal} {\bibinfo
  {journal} {Nucl. Sci. Tech.}\ }\textbf {\bibinfo {volume} {32}},\ \bibinfo
  {pages} {113} (\bibinfo {year} {2021})},\ \Eprint
  {http://arxiv.org/abs/2110.02989} {arXiv:2110.02989 [nucl-th]} \BibitemShut
  {NoStop}%
\bibitem [{\citenamefont {Zheng}\ \emph {et~al.}(2021)\citenamefont {Zheng},
  \citenamefont {Zhang}, \citenamefont {Liu}, \citenamefont {Lin},
  \citenamefont {Shou},\ and\ \citenamefont {Yin}}]{Zheng:2021jrr}%
  \BibitemOpen
  \bibfield  {author} {\bibinfo {author} {\bibfnamefont {L.}~\bibnamefont
  {Zheng}}, \bibinfo {author} {\bibfnamefont {G.-H.}\ \bibnamefont {Zhang}},
  \bibinfo {author} {\bibfnamefont {Y.-F.}\ \bibnamefont {Liu}}, \bibinfo
  {author} {\bibfnamefont {Z.-W.}\ \bibnamefont {Lin}}, \bibinfo {author}
  {\bibfnamefont {Q.-Y.}\ \bibnamefont {Shou}}, \ and\ \bibinfo {author}
  {\bibfnamefont {Z.-B.}\ \bibnamefont {Yin}},\ }\href {\doibase
  10.1140/epjc/s10052-021-09527-5} {\bibfield  {journal} {\bibinfo  {journal}
  {Eur. Phys. J. C}\ }\textbf {\bibinfo {volume} {81}},\ \bibinfo {pages} {755}
  (\bibinfo {year} {2021})},\ \Eprint {http://arxiv.org/abs/2104.05998}
  {arXiv:2104.05998 [hep-ph]} \BibitemShut {NoStop}%
\bibitem [{\citenamefont {Sj\"ostrand}\ \emph {et~al.}(2015)\citenamefont
  {Sj\"ostrand}, \citenamefont {Ask}, \citenamefont {Christiansen},
  \citenamefont {Corke}, \citenamefont {Desai}, \citenamefont {Ilten},
  \citenamefont {Mrenna}, \citenamefont {Prestel}, \citenamefont {Rasmussen},\
  and\ \citenamefont {Skands}}]{Sjostrand:2014zea}%
  \BibitemOpen
  \bibfield  {author} {\bibinfo {author} {\bibfnamefont {T.}~\bibnamefont
  {Sj\"ostrand}}, \bibinfo {author} {\bibfnamefont {S.}~\bibnamefont {Ask}},
  \bibinfo {author} {\bibfnamefont {J.~R.}\ \bibnamefont {Christiansen}},
  \bibinfo {author} {\bibfnamefont {R.}~\bibnamefont {Corke}}, \bibinfo
  {author} {\bibfnamefont {N.}~\bibnamefont {Desai}}, \bibinfo {author}
  {\bibfnamefont {P.}~\bibnamefont {Ilten}}, \bibinfo {author} {\bibfnamefont
  {S.}~\bibnamefont {Mrenna}}, \bibinfo {author} {\bibfnamefont
  {S.}~\bibnamefont {Prestel}}, \bibinfo {author} {\bibfnamefont {C.~O.}\
  \bibnamefont {Rasmussen}}, \ and\ \bibinfo {author} {\bibfnamefont {P.~Z.}\
  \bibnamefont {Skands}},\ }\href {\doibase 10.1016/j.cpc.2015.01.024}
  {\bibfield  {journal} {\bibinfo  {journal} {Comput. Phys. Commun.}\ }\textbf
  {\bibinfo {volume} {191}},\ \bibinfo {pages} {159} (\bibinfo {year}
  {2015})},\ \Eprint {http://arxiv.org/abs/1410.3012} {arXiv:1410.3012
  [hep-ph]} \BibitemShut {NoStop}%
\bibitem [{\citenamefont {Zhang}(1998)}]{Zhang:1997ej}%
  \BibitemOpen
  \bibfield  {author} {\bibinfo {author} {\bibfnamefont {B.}~\bibnamefont
  {Zhang}},\ }\href {\doibase 10.1016/S0010-4655(98)00010-1} {\bibfield
  {journal} {\bibinfo  {journal} {Comput. Phys. Commun.}\ }\textbf {\bibinfo
  {volume} {109}},\ \bibinfo {pages} {193} (\bibinfo {year} {1998})},\ \Eprint
  {http://arxiv.org/abs/nucl-th/9709009} {arXiv:nucl-th/9709009} \BibitemShut
  {NoStop}%
\bibitem [{\citenamefont {Zheng}\ \emph {et~al.}(2024)\citenamefont {Zheng},
  \citenamefont {Liu}, \citenamefont {Lin}, \citenamefont {Shou},\ and\
  \citenamefont {Yin}}]{Zheng:2024xyv}%
  \BibitemOpen
  \bibfield  {author} {\bibinfo {author} {\bibfnamefont {L.}~\bibnamefont
  {Zheng}}, \bibinfo {author} {\bibfnamefont {L.}~\bibnamefont {Liu}}, \bibinfo
  {author} {\bibfnamefont {Z.-W.}\ \bibnamefont {Lin}}, \bibinfo {author}
  {\bibfnamefont {Q.-Y.}\ \bibnamefont {Shou}}, \ and\ \bibinfo {author}
  {\bibfnamefont {Z.-B.}\ \bibnamefont {Yin}},\ }\href {\doibase
  10.1140/epjc/s10052-024-13378-1} {\bibfield  {journal} {\bibinfo  {journal}
  {Eur. Phys. J. C}\ }\textbf {\bibinfo {volume} {84}},\ \bibinfo {pages}
  {1029} (\bibinfo {year} {2024})},\ \Eprint {http://arxiv.org/abs/2404.18829}
  {arXiv:2404.18829 [nucl-th]} \BibitemShut {NoStop}%
\bibitem [{\citenamefont {He}\ and\ \citenamefont {Lin}(2017)}]{He:2017tla}%
  \BibitemOpen
  \bibfield  {author} {\bibinfo {author} {\bibfnamefont {Y.}~\bibnamefont
  {He}}\ and\ \bibinfo {author} {\bibfnamefont {Z.-W.}\ \bibnamefont {Lin}},\
  }\href {\doibase 10.1103/PhysRevC.96.014910} {\bibfield  {journal} {\bibinfo
  {journal} {Phys. Rev. C}\ }\textbf {\bibinfo {volume} {96}},\ \bibinfo
  {pages} {014910} (\bibinfo {year} {2017})},\ \Eprint
  {http://arxiv.org/abs/1703.02673} {arXiv:1703.02673 [nucl-th]} \BibitemShut
  {NoStop}%
\bibitem [{\citenamefont {Shao}\ \emph {et~al.}(2020)\citenamefont {Shao},
  \citenamefont {Chen}, \citenamefont {Ko},\ and\ \citenamefont
  {Lin}}]{Shao:2020sqr}%
  \BibitemOpen
  \bibfield  {author} {\bibinfo {author} {\bibfnamefont {T.}~\bibnamefont
  {Shao}}, \bibinfo {author} {\bibfnamefont {J.}~\bibnamefont {Chen}}, \bibinfo
  {author} {\bibfnamefont {C.~M.}\ \bibnamefont {Ko}}, \ and\ \bibinfo {author}
  {\bibfnamefont {Z.-W.}\ \bibnamefont {Lin}},\ }\href {\doibase
  10.1103/PhysRevC.102.014906} {\bibfield  {journal} {\bibinfo  {journal}
  {Phys. Rev. C}\ }\textbf {\bibinfo {volume} {102}},\ \bibinfo {pages}
  {014906} (\bibinfo {year} {2020})},\ \Eprint
  {http://arxiv.org/abs/2012.10037} {arXiv:2012.10037 [nucl-th]} \BibitemShut
  {NoStop}%
\bibitem [{\citenamefont {Zheng}\ \emph {et~al.}(2020)\citenamefont {Zheng},
  \citenamefont {Zhang}, \citenamefont {Shi},\ and\ \citenamefont
  {Lin}}]{Zheng:2019alz}%
  \BibitemOpen
  \bibfield  {author} {\bibinfo {author} {\bibfnamefont {L.}~\bibnamefont
  {Zheng}}, \bibinfo {author} {\bibfnamefont {C.}~\bibnamefont {Zhang}},
  \bibinfo {author} {\bibfnamefont {S.~S.}\ \bibnamefont {Shi}}, \ and\
  \bibinfo {author} {\bibfnamefont {Z.~W.}\ \bibnamefont {Lin}},\ }\href
  {\doibase 10.1103/PhysRevC.101.034905} {\bibfield  {journal} {\bibinfo
  {journal} {Phys. Rev. C}\ }\textbf {\bibinfo {volume} {101}},\ \bibinfo
  {pages} {034905} (\bibinfo {year} {2020})},\ \Eprint
  {http://arxiv.org/abs/1909.07191} {arXiv:1909.07191 [nucl-th]} \BibitemShut
  {NoStop}%
\bibitem [{\citenamefont {Lin}\ \emph {et~al.}(2005)\citenamefont {Lin},
  \citenamefont {Ko}, \citenamefont {Li}, \citenamefont {Zhang},\ and\
  \citenamefont {Pal}}]{Lin:2004en}%
  \BibitemOpen
  \bibfield  {author} {\bibinfo {author} {\bibfnamefont {Z.-W.}\ \bibnamefont
  {Lin}}, \bibinfo {author} {\bibfnamefont {C.~M.}\ \bibnamefont {Ko}},
  \bibinfo {author} {\bibfnamefont {B.-A.}\ \bibnamefont {Li}}, \bibinfo
  {author} {\bibfnamefont {B.}~\bibnamefont {Zhang}}, \ and\ \bibinfo {author}
  {\bibfnamefont {S.}~\bibnamefont {Pal}},\ }\href {\doibase
  10.1103/PhysRevC.72.064901} {\bibfield  {journal} {\bibinfo  {journal} {Phys.
  Rev. C}\ }\textbf {\bibinfo {volume} {72}},\ \bibinfo {pages} {064901}
  (\bibinfo {year} {2005})},\ \Eprint {http://arxiv.org/abs/nucl-th/0411110}
  {arXiv:nucl-th/0411110} \BibitemShut {NoStop}%
\bibitem [{\citenamefont {Li}\ and\ \citenamefont {Ko}(1995)}]{Li:1995pra}%
  \BibitemOpen
  \bibfield  {author} {\bibinfo {author} {\bibfnamefont {B.-A.}\ \bibnamefont
  {Li}}\ and\ \bibinfo {author} {\bibfnamefont {C.~M.}\ \bibnamefont {Ko}},\
  }\href {\doibase 10.1103/PhysRevC.52.2037} {\bibfield  {journal} {\bibinfo
  {journal} {Phys. Rev. C}\ }\textbf {\bibinfo {volume} {52}},\ \bibinfo
  {pages} {2037} (\bibinfo {year} {1995})},\ \Eprint
  {http://arxiv.org/abs/nucl-th/9505016} {arXiv:nucl-th/9505016} \BibitemShut
  {NoStop}%
\bibitem [{\citenamefont {Christiansen}\ and\ \citenamefont
  {Skands}(2015)}]{Christiansen:2015yqa}%
  \BibitemOpen
  \bibfield  {author} {\bibinfo {author} {\bibfnamefont {J.~R.}\ \bibnamefont
  {Christiansen}}\ and\ \bibinfo {author} {\bibfnamefont {P.~Z.}\ \bibnamefont
  {Skands}},\ }\href {\doibase 10.1007/JHEP08(2015)003} {\bibfield  {journal}
  {\bibinfo  {journal} {JHEP}\ }\textbf {\bibinfo {volume} {08}},\ \bibinfo
  {pages} {003} (\bibinfo {year} {2015})},\ \Eprint
  {http://arxiv.org/abs/1505.01681} {arXiv:1505.01681 [hep-ph]} \BibitemShut
  {NoStop}%
\bibitem [{\citenamefont {Cui}\ \emph {et~al.}(2022)\citenamefont {Cui},
  \citenamefont {Yin},\ and\ \citenamefont {Zheng}}]{Cui:2022puv}%
  \BibitemOpen
  \bibfield  {author} {\bibinfo {author} {\bibfnamefont {P.}~\bibnamefont
  {Cui}}, \bibinfo {author} {\bibfnamefont {Z.}~\bibnamefont {Yin}}, \ and\
  \bibinfo {author} {\bibfnamefont {L.}~\bibnamefont {Zheng}},\ }\href
  {\doibase 10.1140/epja/s10050-022-00709-7} {\bibfield  {journal} {\bibinfo
  {journal} {Eur. Phys. J. A}\ }\textbf {\bibinfo {volume} {58}},\ \bibinfo
  {pages} {53} (\bibinfo {year} {2022})},\ \Eprint
  {http://arxiv.org/abs/2203.13416} {arXiv:2203.13416 [hep-ph]} \BibitemShut
  {NoStop}%
\bibitem [{\citenamefont {Acharya}\ \emph {et~al.}(2021)\citenamefont {Acharya}
  \emph {et~al.}}]{ALICE:2020jsh}%
  \BibitemOpen
  \bibfield  {author} {\bibinfo {author} {\bibfnamefont {S.}~\bibnamefont
  {Acharya}} \emph {et~al.} (\bibinfo {collaboration} {ALICE}),\ }\href
  {\doibase 10.1140/epjc/s10052-020-08690-5} {\bibfield  {journal} {\bibinfo
  {journal} {Eur. Phys. J. C}\ }\textbf {\bibinfo {volume} {81}},\ \bibinfo
  {pages} {256} (\bibinfo {year} {2021})},\ \Eprint
  {http://arxiv.org/abs/2005.11120} {arXiv:2005.11120 [nucl-ex]} \BibitemShut
  {NoStop}%
\bibitem [{\citenamefont {Loizides}\ \emph {et~al.}(2018)\citenamefont
  {Loizides}, \citenamefont {Kamin},\ and\ \citenamefont
  {d'Enterria}}]{Loizides:2017ack}%
  \BibitemOpen
  \bibfield  {author} {\bibinfo {author} {\bibfnamefont {C.}~\bibnamefont
  {Loizides}}, \bibinfo {author} {\bibfnamefont {J.}~\bibnamefont {Kamin}}, \
  and\ \bibinfo {author} {\bibfnamefont {D.}~\bibnamefont {d'Enterria}},\
  }\href {\doibase 10.1103/PhysRevC.97.054910} {\bibfield  {journal} {\bibinfo
  {journal} {Phys. Rev. C}\ }\textbf {\bibinfo {volume} {97}},\ \bibinfo
  {pages} {054910} (\bibinfo {year} {2018})},\ \bibinfo {note} {[Erratum:
  Phys.Rev.C 99, 019901 (2019)]},\ \Eprint {http://arxiv.org/abs/1710.07098}
  {arXiv:1710.07098 [nucl-ex]} \BibitemShut {NoStop}%
\bibitem [{\citenamefont {Dai}\ \emph {et~al.}(2024)\citenamefont {Dai},
  \citenamefont {Zhao},\ and\ \citenamefont {He}}]{Dai:2024vjy}%
  \BibitemOpen
  \bibfield  {author} {\bibinfo {author} {\bibfnamefont {Y.}~\bibnamefont
  {Dai}}, \bibinfo {author} {\bibfnamefont {S.}~\bibnamefont {Zhao}}, \ and\
  \bibinfo {author} {\bibfnamefont {M.}~\bibnamefont {He}},\ }\href@noop {} {\
  (\bibinfo {year} {2024})},\ \Eprint {http://arxiv.org/abs/2402.03692}
  {arXiv:2402.03692 [hep-ph]} \BibitemShut {NoStop}%
\end{thebibliography}%

\end{document}